\documentclass[11pt]{article}
\usepackage{CJK}
\usepackage{graphicx}
\usepackage{fancyhdr}
\usepackage{amsmath,amsthm,amssymb}
\usepackage{graphicx,color,epstopdf}
\usepackage{mathrsfs}
\usepackage{multirow}
\usepackage{indentfirst}
\usepackage{color}
\renewcommand{\baselinestretch}{1.15}
\newtheorem{theorem}{THEOREM}
\newtheorem{remark}{REMARK}
\newtheorem{lemma}{LEMMA}

\makeatletter
\newcommand{\Rmnum}[1]{\expandafter\@slowromancap\romannumeral #1@}
\makeatother
\newtheorem{corollary}{COROLLARY}
\allowdisplaybreaks

\begin{document}
\date{}

\title
{\Large \bf Enhancements of nonparametric generalized
likelihood ratio test: Bias-correction and dimension reduction } {\small
\author{Cuizhen Niu$^1$, Xu Guo$^2$ and Lixing Zhu$^3$ \footnote{Corresponding author: Lixing Zhu, email: lzhu@hkbu.edu.hk. The research described herewith was supported by a grant from the University Grants Council of Hong Kong, China and also supported by the Outstanding Innovative Talents
Cultivation Funded Programs 2014 of Renmin University of China.}\\
{\small {\small {\it $^1$School of Statistics, Renmin
University of China, Beijing } }}\\
{\small {\small {\it $^2$ Nanjing University of Aeronautics and Astronautics, Nanjing
} }} \\
{\small {\small {\it $^3$Department of Mathematics, Hong Kong Baptist
University, Hong Kong } }}
 }}
\date{}
\maketitle

\date{}
\maketitle

\renewcommand\baselinestretch{1.4}
{\small

\noindent {\bf Abstract:} Nonparametric generalized likelihood ratio test is  popularly used for model checking for regressions. However, there are two issues that may be the barriers for its powerfulness. First,  the bias term in its liming null distribution  causes the test not to  well control type I error and thus  Monte Carlo approximation for critical value determination is required. Second, it  severely suffers from the curse of dimensionality due to the use of multivariate nonparametric function estimation.  The purpose of this paper is thus two-fold: a bias-correction is suggested to this test and a dimension reduction-based model-adaptive enhancement is recommended to promote the power performance. The proposed test still possesses the Wilks phenomenon, and  the test statistic can converge to its limit at a  much faster rate and is much more sensitive to alternative models than the original nonparametric generalized likelihood ratio test as if the dimension of covariates were one.  Simulation studies are conducted to evaluate the finite sample performance and to compare with other popularly used tests. A real data analysis is conducted  for illustration.

\bigskip

 }
\newpage

\section{Introduction}
\renewcommand{\theequation}{1.\arabic{equation}}
\setcounter{equation}{0}
Parametric regression models, which take specifically parametric forms for the regression relationship
between response and predictors, are commonly used and studied in practice. This is due to their well established theories, easy implementation and interpretation. However, when the dimension of
the predictor vector is high, even moderate, it is difficult to correctly specify the regression form. People also concern whether a specifically parametric regression  can fit the data adequately. Therefore, it is
necessary to conduct model checking to determine a suitable regression model  before any further statistical analysis. To avoid model mis-specification, nonparametric regression models are proposed and investigated. However, under this case,  the nonparametric estimation is usually inaccurate, which is documented as `curse of
dimensionality'.

Consider the following parametric single-index model
\begin{equation}\label{null_model}
  Y=g(\beta^\top X,\theta)+e,
\end{equation}
where $Y$ is the response with the covariate $X\in \mathbb{R}^p$,
$\beta$ and $\theta$ are the parameter vectors of dimensions $p$ and
$d$, respectively and $E(e|X)=0$. Besides, $g(\cdot)$ is a known
link function and the superscript $\top$ denotes transposition. The
nonparametric regression model takes the following form:
\begin{equation}\label{nonpara_model}
    Y=m(X)+\varepsilon,
\end{equation}
where $m(\cdot)$ is the unknown mean function and
$E(\varepsilon|X)=0$. There exist several proposals available in the literature to
test  model (\ref{null_model}) against model
(\ref{nonpara_model}). To the best of our knowledge, all of existing tests can usually be classified
into two categories: local smoothing methods and global smoothing
methods with their respective advantages and
disadvantages. The former generally relies on nonparametric regression
estimators and the latter class involves  empirical processes. As
commented in Guo et al. (2014), the former is more sensitive to
high-frequency alternative models while the latter may be in favor
of smooth alternatives. On the other hand, it is well known that the former severely suffers from curse of
dimensionality since the typical convergence rate is only
$O(n^{-1/2}h^{-p/4})$, which is very slow in high-dimensional
situation. Due to the data sparseness in high-dimensional space, the
power performance of the latter often drops down very significantly. Further,
for these classical tests, compute-intensive re-sampling technique
or Monte Carlo approximation are usually needed to help
critical value determination or $p$-value computation.

Varieties of methods can be included into the first class. Among others,
the quadratic conditional moment-based test was proposed by Zheng (1996),
the minimum distance test developed by Koul and Ni (2004) and the
distribution distance test introduced by Van Keilegom et al (2008).
See also H\"{a}rdle and Mammen (1993), Dette (1999) and Zhang and
Dette (2004). Turn to the latter category. 
A test that is based on residual-marked empirical process was presented by Stute (1997).  An innovation approach for the above empirical process was suggested by Stute et. al (1998b). A relevant reference for
generalized linear models is Stute and Zhu (2002). Koul and Stute
(1999) studied a class of tests for an autoregressive model, which
are based on empirical processes marked by certain residuals. Stute
et. al (1998a) recommended the wild bootstrap to approximate the limiting null
distribution of the empirical process-based test statistic. For a comprehensive
review, see  Gonz$\acute{a}$lez-Manteiga and Crujeiras
(2013).

Inspired by the classical likelihood ratio test, Fan et al. (2001)
introduced a nonparametric generalized likelihood ratio (NGLR) test for the above hypothetical model in (\ref{null_model}) versus the alternative model in (\ref{nonpara_model}) . A
significant property of  NGLR is that the limiting null distribution
does not depend on nuisance functions, exhibiting what is known as
Wilks phenomenon. This test has been applied in many different
situations such as the varying-coefficient partially linear models
(Fan and Huang, 2005) and partially linear additive models (Jiang et. al, 2007). For more details, see Fan and Jiang
(2007). However, similar to other local smoothing tests, NGLR also
suffers from the curse of dimensionality because of the inevitable use of
multivariate nonparametric function estimation. To be precise, under
the null hypothesis, the NGLR test statistic converges to its limit
at the rate of order $O(n^{-1/2}h^{-p/4})$ and  can only
detect alternatives distinct from the null hypothesis at the rate of
order $O(n^{-1/2}h^{-p/4})$. Further, its limit has a bias term and thus, as Fan and Li (1997) and the later research work in this field pointed out, the bias term causes the great difficulty of controlling type I error. Also,  as Dette and von Lieres und Wilkau (2001) pointed out, for finite sample sizes, the bias converging to infinity has to be taken into account. See also Zhang and Dette (2004).
  To determine critical values, resampling/Monte Carlo approximation such as the bootstrap method is required even the Wilk's phenomenon holds. But the
bootstrap approximation is  computationally intensive and  even though type I error can not well be  controlled
 when the dimension of predictors is high. We will see this in the simulation section below. Second, when the dimension $p$ is even moderate
the power can be very low. In the simulation section, we will show that the power drops down significantly with increasing the dimension. Obviously, it is desirable to have a test that can well control type I error  without the assistance of resampling approximation  and more importantly, can handle high-dimensional data with acceptable performance.

Note that for the parametric single-index model (\ref{null_model}),
the regression relationship of the response $Y$ depends on the
predictors $X$ only through the one-dimensional index $\beta^\top
X$. In other words,  $\beta^\top X$ can
capture all the regression information on the response. Thus, the model
(\ref{null_model}) can be viewed as a model with a dimension-reduction structure.
Motivated by this feature, Stute and Zhu (2002) proposed a
 test that is fully based on $\beta^\top X$ for the  model
(\ref{null_model}) as follows:
\begin{equation*}
  R_n(u)=\frac{1}{\sqrt n}\sum_{i=1}^n I_{\{\hat\beta^\top x_i\leq u\}}[y_i-m(\hat\beta^\top x_i,\hat\theta)],
\end{equation*}
where $\hat\beta$ and $\hat\theta$ are, under the null hypothesis, two respective
root-$n$ consistent estimates of $\beta$ and $\theta$,
and $I(\cdot)$ is the indicator function. The above test
statistic can greatly avoid the dimensionality. But it is obviously a
directional test that cannot detect general alternatives.  Guo et al.
(2015) discussed this problem in details and proposed a test that has the nature of dimension reduction and still is omnibus. 



In this paper, we try to simultaneously solve these two problems from which  the original NGLR suffers by reducing both the bias dimensionality. To this end, we will propose a bias-correction based version of NGLR such that the bias can asymptotically be removed and   combine the idea in  Guo et al. (2015) with NGLR to  circumvent the
curse of dimensionality. The details are as follows.

Consider a more
 general alternative model:
\begin{eqnarray}\label{gen-alter1}
Y=m(B^{\top}{X})+\varepsilon,
\end{eqnarray}
where $B$ is a $p\times q$ orthonormal matrix with $B^{\top}B=I_q$ being a $q$ dimensional identity matrix where $q$ is unknown with $1\le q\le p$, $m(\cdot)$ is an unknown
smooth function and $E(\varepsilon|{X})=0$.  Under the null
hypothesis, the model (\ref{gen-alter1}) reduces to the model
(\ref{null_model}) with $q=1,B=\beta/||\beta||_2$ and
$m(\cdot)=g(\cdot)$. Here $\|\cdot\|_2$ denotes the $L_2$ norm.
Under the alternative hypothesis, the model (\ref{gen-alter1}) can
be regarded as a multi-index model with $q\geq 1$ indices.
If the alternative model is a purely nonparametric
regression model as (\ref{nonpara_model}), $q$ is equal to $p$.

 As we found that the bias comes from the slow convergence rate of nonparametric estimation in the residual sum of squares, we then consider a  sum of residual product that are from parametric and nonparametric fits. It will be confirmed theoretically that the bias can be asymptotically removed. As a by-product,  the bias-reduction-based NGLR statistic also has a smaller asymptotic variance.  Through the simulation studies later, we can see its advantages on significance level maintainance and power enhancement.

 As for dimensionality reduction, the key  is how to adapt the model structure under the null hypothesis and under the alternative how the test can automatically adapt the model structure such that the test is omnibus. To achieve this goal, we need to automatically identify $\beta$ and $B$, and then construct a test that is based on this model-adaptive identification to  fully use the information provided by the underlying model.
The estimation procedure will be described in the next section. the  model-adaptive version of NGLR does have two desired features: the Wilks
phenomenon still holds and the test statistic can converge to the
limit at the rate of order $O(n^{-1/2}h^{-1/4})$
under the null hypothesis and can detect local alternatives distinct
from the null at the rate of order $O(n^{-1/2}h^{-1/4})$ rather than $O(n^{-1/2}h^{-p/4})$.

The rest of this article is organized as follows. In Section~2, the model-adaptive enhancement versions of the NGLR tests without bias-correction and with bias-correction are both constructed. The methods to
estimate $B$ and identify  the structure dimension $q$ are also presented
in this section. The asymptotic properties under the null hypothesis, local and global alternative hypothesis are investigated in Section~3. Simulation studies and a real data analysis are conducted to evaluate the finite sample performance in Sections~4 and ~5, respectively. All the technical proofs are
relegated to  Appendix.

%
%
%
%
%
%
%
%

\section{Test statistic construction and relevant properties}\label{sec2}
\renewcommand{\theequation}{2.\arabic{equation}}
\setcounter{equation}{0}

The hypotheses are:
\begin{eqnarray*}
    &&H_0: E(Y|X)=g(\beta^\top X,\theta)\,\, \mbox{for some}\,\, \beta\in\mathbb{R}^p,\theta\in\mathbb{R}^d\,\,;\,\,\nonumber \\
    &&H_1: E(Y|X)=m(B^\top X)\neq g(\beta^\top X,\theta), \,\, \mbox{for any}\,\, \beta\in\mathbb{R}^p,\theta\in\mathbb{R}^d,
\end{eqnarray*}

\subsection{Model-adaptive enhancement of the NGLR test}
  Suppose $(y_i,x_i),\,\,i=1,\ldots,n$ is a random sample and the
error $\varepsilon_i$ in nonparametric model (\ref{nonpara_model})
is i.i.d $N(0,\sigma^2)$. The normality is just used for motivating the test statistic construction. As Fan et al (2001) did, the conditional log-likelihood function of $y_i$ given
$x_i$ can be written as:
\begin{equation*}
   \hat l(m,\sigma^2)=-\frac{n}{2}\ln(2\pi\sigma^2)-\frac{1}{2\sigma^2}\sum_{i=1}^n [y_i-m(x_i)]^2.
\end{equation*}
Under the null hypothesis $H_0$,
\begin{equation}\label{loglike_H0}
    \hat
    l(\hat\beta,\hat\theta,\sigma^2)=-\frac{n}{2}\ln(2\pi\sigma^2)-\frac{1}{2\sigma^2}\mbox{RSS}_0.
\end{equation}
Here $\mbox{RSS}_0$ is the residual sum of squares under the hypothetical
model (\ref{null_model}), that is,
\begin{equation}\label{RSS0}
    \mbox{RSS}_0=\sum_{i=1}^n [y_i-g(\hat\beta^\top x_i,\hat\theta)]^2,
\end{equation}
where $\hat\beta$ and $\hat\theta$ denote the ordinary least squares (OLS) estimates of the parameters $\beta$ and $\theta$, respectively.
Further, maximizing the likelihood in (\ref{loglike_H0}) with
respect to nuisance parameter $\sigma^2$ yields
$\hat\sigma^2=n^{-1}\mbox{RSS}_0$ and substituting the estimate in
(\ref{loglike_H0}) yields the following log-likelihood function:
\begin{equation}\label{newloglike_H0}
    \hat l(\hat\beta,\hat\theta,\hat\sigma^2)=-\frac{n}{2}\ln(\mbox{RSS}_0)-\frac{n}{2}[
    1+\ln(2\pi/n)].
\end{equation}
Under the alternative hypothesis $H_1$, with a similar argument, we
can obtain
\begin{equation}\label{newloglike_H1}
    \hat l(\hat m, \hat B(\hat q),\hat\sigma^2)=-\frac{n}{2}\ln(\mbox{RSS}_1)-\frac{n}{2}[
    1+\ln(2\pi/n)].
\end{equation}
Here $\hat B(\hat q)$ is an estimate of $B$. $\mbox{RSS}_1$ is the residual sum of squares under the
alternative model (\ref{gen-alter1}), that is,
\begin{equation}\label{RSS1_bias}
    \mbox{RSS}_1=\sum_{i=1}^n [y_i-\hat m(\hat B(\hat q)^\top x_i)]^2.
\end{equation}
We will later specify the estimate $\hat m(\hat B(\hat q)^\top x_i)$ of $m(B^\top x_i)$.

Under the null hypothesis $H_0$, $\hat l(\hat m, \hat B(\hat
q),\hat\sigma^2)$ should be  close to $\hat
l(\hat\beta,\hat\theta,\hat\sigma^2)$. While under the alternative
hypothesis $H_1$, they should deviate from each other. This motivates us to
define the following  test statistic:
%
\begin{equation}\label{statistic}
    T_n=\hat l(\hat m, \hat B(\hat q),\hat\sigma^2)-\hat l(\hat\beta,\hat\theta,\hat\sigma^2)=
    \frac{n}{2}\log\frac{\mbox{RSS}_0}{\mbox{RSS}_1}
    \approx  \frac{n}{2}\frac{(RSS_0-RSS_1)}{RSS_1}.
\end{equation}

From the above construction, it seems there is no difference from the original NGLR test except having an extra estimate of $B$. However, we will see that using an extra estimate $\hat B(\hat
q)$ of $B$ will play a very important role in the model-adaption of the test.

Once an estimate $\hat B(\hat q)$ of $
B$ is available, $m(B^\top x_i)$ can be estimated by:
\begin{equation}\label{hatm}
  \hat m(\hat B(\hat q)^\top x_i)=\sum_{j=1}^n w_{ij}(\hat B(\hat q)) y_j.
\end{equation}
Here
\begin{equation*}
  w_{ij}(\hat B(\hat q))=\frac{\mathcal{K}\{\hat B(\hat q)^\top(x_i-x_j)/h\}}{\sum_{l=1}^n \mathcal{K}\{\hat B(\hat q)^\top(x_i-x_l)/h\}},
\end{equation*}
with $\mathcal{K} (\cdot)$ being a $\hat q$-dimensional kernel function and
$h$ being the bandwidth. Moreover $\mathcal{K} (\cdot)$ is a symmetric
kernel with compact support, and is of order $r\geq2$, that is,
\begin{equation*}
  \frac{(-1)^r}{r!}\int u^j \mathcal{K}(u)du=
  \begin{cases}
    1 & \text{if $j=0$,}\\
    0 & \text{if $1\leq j\leq r-1$,}\\
    k_r>0 & \text{if $j=r$.}
  \end{cases}
\end{equation*}
%
%

Now we can make the comparison with NGLR. The main difference is the definition of $\mbox{RSS}_1$. In
Fan at al. (2001),
\begin{equation*}
    \overline{\mbox{RSS}}_1=\sum_{i=1}^n [y_i-\hat m(x_i)]^2.
\end{equation*}
Here $\hat m(x_i)=\sum_{j=1}^n \tilde K_h({x}_i-{x}_j)
y_j/\sum_{j=1}^n \tilde K_h({x}_i-{x}_j)$, $\tilde
K_h(\cdot)=\tilde{K}(\cdot/h)/h^p$, with $\tilde{K}(\cdot)$ a
$p$-dimensional kernel function. Then  NGLR is defined as
\begin{equation}\label{fanstatis}
    \overline{T}_n=    \frac{n}{2}\log\frac{\mbox{RSS}_0}{\overline{\mbox{RSS}}_1}
    \approx  \frac{n}{2}\frac{(RSS_0-\overline{\mbox{RSS}}_1)}{\overline{\mbox{RSS}}_1}.
\end{equation}

We will see this difference will make the test behave very different from the original NGLR.
%

\subsection{Bias-corrected version of the test statistic}

From  Theorem~\ref{theo1} in Section~3 below, it can be seen
clearly that there exists an asymptotic bias converging to infinity in the limiting null
distribution, which has to be taken into account in practice.
In this subsection, we aim to propose a bias-correction method to remove
the asymptotic bias. The motivation is from the theoretical investigation for the test statistic. The asymptotic bias of $T_n$ in (\ref{statistic}) is caused by the slow convergence rate of
$\hat m(\hat B(\hat q)^\top x_i)$ to $m(B^\top x_i)$ in $RSS_1$. To eliminate the bias asymptotically, we replace
$\mbox{RSS}_1$ in (\ref{RSS1_bias}) by
\begin{equation}\label{RSS1_nobias}
    \widetilde{\mbox{RSS}}_1=\sum_{i=1}^n \Big|[y_i-\hat {\tilde m}(\hat B(\hat q)^\top x_i)][y_i-g(\hat\beta^\top x_i,\hat\theta)]\Big|,
\end{equation}
where, $|\cdot|$ denotes the absolute value and
the leave-one-out kernel estimate of $m(B^\top x_i)$ is applied. To be precise,
\begin{equation}\label{wanwanw}
  \hat {\tilde m}(\hat B(\hat q)^\top x_i)=\frac{\sum_{j\neq i}^n\mathcal{K}\{\hat B(\hat q)^\top(x_i-x_j)/h\}y_j}{\sum_{l\neq i}^n \mathcal{K}\{\hat B(\hat q)^\top(x_i-x_l)/h\}}
  =:\sum_{j\neq i}^n \tilde{w}_{ij}(\hat B(\hat q))y_j,
\end{equation}
which is different from the original kernel estimate $\hat m(\hat B(\hat q)^\top x_i)$ in (\ref{hatm}).
It is clear that the term $\widetilde{\mbox{RSS}}_1$ represents the sum
of residual product  under the null and alternative hypothesis.
The bias-corrected version of the test statistic $T_n$ in (\ref{statistic})
is defined as:
\begin{eqnarray}
  \tilde{T}_n&=&\frac{n}{2}
    \frac{\mbox{RSS}_0-\widetilde{\mbox{RSS}}_1}{\widetilde{\mbox{RSS}}_1}\nonumber\\
    &=&\frac{n}{2}\frac{\sum_{i=1}^n |\hat e_i|\big\{|\hat e_i|-|y_i-\hat {\tilde m}(\hat B(\hat q)^\top x_i)|\big\}}{\sum_{i=1}^n \Big|\hat e_i[y_i-\hat {\tilde m}(\hat B(\hat q)^\top x_i)]\Big|}~~~~~~~~~\label{nobias_statistic}
\end{eqnarray}
where $\hat e_i=y_i-g(\hat\beta^\top x_i,\hat\theta).$

To understand the difference between NGLR and the bias-corrected NGLR under the null hypothesis, we can check mainly the difference between two numerators of the two test statistics because the denominator goes to a constant in probability. Recall that  the main reason to have the bias term is the slow convergence rate of the nonparametric estimation in $\mbox{RSS}_1$ to the regression function under the null hypothesis. The parametric fitting in $\widetilde{\mbox{RSS}}_1$ has, under the null hypothesis, a faster convergence rate to the error than the nonparametric fitting, and then reduces the bias asymptotically and at the same time, the another nonparametric fitting can still help   the test  detect the alternative.

%
%

 Bias correction was also investigated in Gozalo
and Linton (2001) for additivity test motivated by Lagrange
multiplier tests. However, there is no  investigation for model checking. Further, unlike
Gozalo and Linton (2001), a leave-one-out kernel estimate is applied which is proved to be useful since without this, the asymptotic
bias still exists although it can be  reduced. Further, we also use absolute summands in $\widetilde{\mbox{RSS}}_1$ to avoid possible negative values.

\subsection{Identification and estimation of $B$}
 Note that the model (\ref{gen-alter1}) is
a multi-index model.
Here, we first estimate the matrix $B$ under the given $q$ and then
study how to select $q$ consistently. To this aim, outer product of
gradients (OPG) and minimum average variance estimation (MAVE)
introduced by Xia et al (2002) are adopted. OPG is easy to implement
and MAVE possesses excellent performance in general.  Review
these two methods below.

\subsubsection{Outer product gradients}
Denote $m(B^\top x)=E(Y|X=x), \nabla
m(B^\top X)=\frac{\partial}{\partial{X}}{m(B^\top X)}$ and
$m'(B^\top X)=\frac{\partial}{\partial{B^\top X}}{m(B^\top X)}$.
Note that
\begin{eqnarray*}
E\{\nabla m(B^\top X)\nabla m(B^\top X)^\top\} = B E\{m'(B^\top
X)m'(B^\top X)^\top\}B^\top
\end{eqnarray*}
owns $q$ nonzero eigenvalues. Therefore, $B$ is in the space spanned
by the  $q$  eigenvectors of $E\{\nabla m(B^\top X)\nabla
m(B^\top X)^\top\}$ corresponding to the largest $q$ eigenvalues.
Through the above analysis, the estimate of $B$ can be obtained by
estimating $E\{\nabla m(B^\top X)\nabla m(B^\top X)^\top\}$.

To implement the estimation, we first estimate the gradients by the local
linear smoother:
\begin{equation*}
  m(B^\top x_i)=m(B^\top x_j)+m'(B^\top x_j)^\top B^\top (x_i-x_j)=a_j+b_j^\top x_{ij},
\end{equation*}
where $a_j=m(B^\top x_j)$, $b_j=B\times m'(B^\top x_j)$ and
$x_{ij}=x_i-x_j$. Then the estimate $(\hat a_j,\hat b_j)$ can be
obtained by solving the following minimization problem:
\begin{equation*}
  \min_{a_j,b_j}\sum_{i=1}^n \mathcal{K}_h(B^\top x_{ij})\{y_i-a_j-b_j^\top x_{ij}\}^2,
\end{equation*}
where $\mathcal{K}_h(\cdot)=\mathcal{K}(\cdot/h)/h$ with
$\mathcal{K}(\cdot)$ being a $q$-dimensional kernel function and $h$
being a bandwidth. The corresponding estimating equation can be
rewritten as
\begin{equation*}
  \sum_{i=1}^n \mathcal{K}_h(B^\top x_{ij})(1, x_{ij}^\top)^\top\{y_i-\hat a_j-\hat b_j^\top x_{ij}\}=0.
\end{equation*}
The estimate of $E\{\nabla m(B^\top X)\nabla m(B^\top
X)^\top\}$ can be constructed as:
\begin{equation}\label{Sigma_OPG}
  \hat\Sigma=\frac{1}{n}\sum_{j=1}^n \hat b_j \hat b_j^\top.
\end{equation}
Thus, the  $q$ eigenvectors that are associated with the largest $q$ eigenvalues of $\hat\Sigma$ can be regarded as the estimate of the matrix $B$.

\subsubsection{Minimum average variance estimation}
The minimum average variance estimation (MAVE) method was first
proposed by Xia et al. (2002). This method needs no strong
assumptions on the probabilistic structure on $X$. 
At the population level, MAVE minimizes the objective function
\begin{equation*}
E\{Y - E(Y|B^\top X)\}^2=E\left(E[\{Y - E(Y|B^\top X)\}^2|B^\top
X]\right)\quad {\rm{subject\ to}}\ B^\top B = I_q.
\end{equation*}

This motivates an estimation procedure for $\sigma_{B}^2(B^\top X)=E[\{Y - E(Y|B^\top
X)\}^2|B^\top X])$.  By using the local linear smoother for $E(Y|B^\top
X)$, $E[\{Y - E(Y|B^\top X)\}^2|B^\top X=B^\top x_j])$ can be
estimated by $\sum_{i=1}^n (y_i-a_j-d_j^\top B^\top x_{ij})^2
\mathcal{K}_h(B^\top x_{ij})$. Finally take the average over all $j$ to get the estimated
objective function.

In sum, when a sample $\{({x}_1, y_1), \ldots, ({x}_n, y_n)\}$ is
available, we can estimate $B$ through minimizing
\begin{equation*}
  \sum_{j=1}^n\sum_{i=1}^n (y_i-a_j-d_j^\top B^\top x_{ij})^2 \mathcal{K}_h(B^\top x_{ij})
\end{equation*}
over all $B$ satisfying $B^\top B = I_q$, $a_j$ and $d_j=m'(B^\top
x_j)$.
 The details of the
algorithm can be found in Xia et al. (2002).
%
%

\subsection{Estimation of structural dimension $q$}
In the above subsection, the structural dimension $q$ is assumed to be known. Now we introduce
two techniques to estimate it:
ridge-type ratio estimate (RRE) and Bayesian information criterion type estimate (BIC) that is motivated by Zhu et al (2006) and Wang and Yin (2008).

{\it Ridge-type Ratio Estimate(RRE)}. This estimate is in spirit similar to the one proposed by
Xia et al (2014), but with a different ridge value. It is very simple and easy to implement. It is the
eigenvalues' ratio modified by adding a positive ridge value $c$.
Determine $\hat q$ by
\begin{equation*}
  \hat q=\arg\min_{k=1,2,\ldots,{p-1}}\frac{\lambda_{k+1}+c}{\lambda_{k}+c},
\end{equation*}
where $\lambda_1\geq\ldots\geq\lambda_p$ are the eigenvalues of
$\hat\Sigma$ in (\ref{Sigma_OPG}). The consistency of $\hat q$ will be proved in the following and  the constant
$c=1/\sqrt{nh}$ is  recommended.

{\it BIC estimate.} For MAVE, it is related to the residual sum
of squares. The above simple ridge-type ratio estimate can not be
used to determine $q$ for the MAVE based estimate. Instead, we use the
modified BIC developed by Wang and Yin (2008):
\begin{equation*}
  \hat q=\arg\min_{k=1,2,\ldots,p} BIC_k=\arg\min_{k=1,2,\ldots,p} \log\Big(\frac{RSS_k}{n}\Big)+\frac{\log(n)k}{\min(nh^k,\sqrt n)},
\end{equation*}
where $RSS_k$ is the residual sum of squares and $k$ is the estimate of the dimension:
\begin{equation*}
  RSS_k= \sum_{j=1}^n\sum_{i=1}^n (y_i-a_j-d_j^\top \hat B(k)^\top x_{ij})^2 \mathcal{K}_h(\hat B(k)^\top x_{ij}).
\end{equation*}
Under some mild conditions, the consistency of $\hat q$ has been
shown by Theorem~1 in Wang and Yin (2008).

\begin{remark}\label{consistency_q}
The following consistency of $\hat q$ is established under both the null and global alternative hypothesis. Under Conditions (C5) and (C7) in Appendix, with a probability going to one, the  ridge-type ratio estimate $\hat q= q$ as $n\rightarrow\infty$. Under the regularity conditions designed in Wang and Yin (2008), the MAVE-based estimate $\hat q= q$ as $n\rightarrow\infty$. Therefore, for a $q\times q$ orthogonal matrix $C$, $\hat B(\hat q)$ is a consistent estimate of $BC^T$. We will show $\hat q= 1$ under the local alternative hypothesis in Section~{\ref{sec3}}.
\end{remark}

\section{Theoretical results}\label{sec3}
\renewcommand{\theequation}{3.\arabic{equation}}
\setcounter{equation}{0}

In this section, the limiting null distributions of our proposed
model-adaptive enhancement  NGLR test $T_n$ in (\ref{statistic}) and the bias-correction version $\tilde{T}_n$ in (\ref{nobias_statistic}) are both derived and their asymptotic
properties under local and fixed alternative hypothesis are also
investigated.

\subsection{Limiting null distribution}

Let ${Z}=\beta^\top X$ and Define
\begin{eqnarray}
  Q_1&=&\Big\{\mathcal{K}(0)-\frac{\int\mathcal{K}^2(u)du}{2}\Big\}\frac{\int\sigma^2(z)dz}{
  \int\sigma^2(z)f(z)dz},\label{theo1_Q}\\
  \eta_0^2&=&2\int\sigma^4(z)dz\int[2\mathcal{K}(u)-
   \mathcal{K}*\mathcal{K}(u)]^2du\label{eta02},\\
  V_0&=&\frac{\eta_0^2}{4(\int\sigma^2(z)f(z)dz)^2},\label{theo1_V}
\end{eqnarray}
where the symbol $*$ denotes the convolution operator and
$K*K(x)=\int K(t)K(x-t)dt$. Besides,
$\sigma^2(z)=E(\varepsilon^2|Z=z)$,
$\sigma^4(z)=[E(\varepsilon^2|Z=z)]^2$ and $f(\cdot)$ is the density
function of $Z$. In order to obtain the estimates $\hat Q_1$ and
$\hat V_0$ for $Q_1$ and $V_0$, we first list the consistent
estimates for the quantities $L_1:=\int \sigma^2(z)f(z)dz$,
$L_2:=\int \mathcal{K}^2(u)du\int \sigma^2(z)dz$,
$L_3:=\mathcal{K}(0)\int \sigma^2(z)dz$ and $\eta_0^2$ as follows:
\begin{eqnarray*}
  \hat L_1&=&\frac{1}{n}\sum_{i=1}^n\xi_i^2=\frac{1}{n}\sum_{i=1}^n[y_i-\hat m(\hat B(\hat q)^\top x_i)]^2,\\
  \hat L_2&=&h\sum_{i=1}^n\sum_{j=1}^n w_{ij}^2(\hat B(\hat q))[y_j-\hat m(\hat B(\hat q)^\top x_j)]^2,\\
  \hat L_3&=&h\sum_{i=1}^n w_{ii}(\hat B(\hat q))[y_i-\hat m(\hat B(\hat q)^\top x_i)]^2,\\
  \hat\eta_0^2&=&2h\sum_{i=1}^n\sum_{i\neq j}^n[y_i-\hat m(\hat B(\hat q)^\top x_i)]^2[y_j-\hat m(\hat B(\hat q)^\top x_j)]^2\\
  &&\times\Big\{
  w_{ij}(\hat B(\hat q))+w_{ji}(\hat B(\hat q))-\sum_{k=1}^nw_{ki}(\hat B(\hat q))w_{kj}(\hat B(\hat q))\Big\}^2.
\end{eqnarray*}
Therefore, we have
\begin{equation*}
  \hat Q_1=\frac{1}{\hat L_1}(\hat L_3-\frac{\hat L_2}{2}),\qquad
  \hat V_0=\frac{\hat\eta_0^2}{4\hat L_1^2}.
\end{equation*}

The following theorem states the limiting null distribution of the
proposed test statistic $T_n$ in (\ref{statistic}).

\begin{theorem}\label{theo1}
Under the null hypothesis in (\ref{null_model}) and conditions (C1)-(C7)
in  Appendix, we have:
\begin{equation*}
  \sqrt{h}\Big(T_n-\frac{Q_1}{h}\Big)
  \Rightarrow N(0,V_0).
\end{equation*}
\end{theorem}
Plugging in their consistent estimates and by Slutsky's Theorem, the
following corollary can be easily obtained.

\begin{corollary}\label{coro1}
Under the null hypothesis in (\ref{null_model}) and conditions (C1)-(C7)
in  Appendix, we have:
\begin{equation}\label{formulaSn}
S_n:=\frac{\sqrt h}{\sqrt {\hat
V_0}}\Big (T_n-\frac{\hat Q_1}{h}\Big )\Rightarrow N(0, 1).
\end{equation}

\end{corollary}

Theorem~\ref{theo1} and Corollary~\ref{coro1} characterize the
asymptotic normality of the  proposed test statistic. The null hypothesis $H_0$  is rejected when $|S_n|\ge Z_{1-\alpha/2}$ that is the $1-\alpha/2$ upper quantile of the distribution $N(0,1)$.  It is
notable that with homoscedastic errors, the asymptotic bias and
variance in Theorem~\ref{theo1} is free of any nuisance
parameters and nuisance functions. To be precise, $Q_1=
|\Omega|\cdot\{\mathcal{K}(0)-\int \mathcal{K}^2(u)du/2\}$ and
$V_0=|\Omega/2|\int [2\mathcal{K}(u)-\mathcal{K}*\mathcal{K}(u)]^2
du$, here $|\Omega|$ denotes the Lebesgue's measure of the support
$Z$. Thus, similar to NGLR, the limiting null
distribution of the proposed test statistic is free of nuisance
parameters enjoying Wilks phenomenon. Note that
$n^{-1}(RSS_0-{RSS}_1)=O_p(\sqrt h)$ under $H_0$. Thus
Theorem~\ref{theo1} and Corollary~\ref{coro1} show that $n^{-1}(RSS_0-{RSS}_1)=O_p(\sqrt h)$ under $H_0$ and the
asymptotic bias of $T_n$ is at the order of $h^{-1/2}$. However,  for NGLR, $n^{-1}(RSS_0-{RSS}_1)=O_p( h^{p/2})$ and the asymptotic bias has the order of $h^{-p/2}$. Thus, the convergence rate is greatly
improved, the dimensionality effect is significantly eliminated and the bias is much reduced.  These results make that the proposed test
controls the size better compared with the  NGLR test. The finite sample simulations later also confirm this claim.

\vskip 10pt

We then state the asymptotic property of the bias-correction version test statistic $\tilde{T}_n$ in (\ref{nobias_statistic}) under the null hypothesis.
\begin{theorem}\label{theo_nullnobias}
Under the same conditions as those in Theorem~\ref{theo1},
we have
\begin{equation*}
  \sqrt{h}\tilde{T}_n\Rightarrow N(0,V_1),
\end{equation*}
where
\begin{eqnarray}\label{V1}
  V_1&=&\frac{\int\sigma^4(z)dz\int K^2(u)du}{2(\int\sigma^2(z)f(z)dz)^2}.
\end{eqnarray}
\end{theorem}

Compared with Theorem~\ref{theo1}, the significant difference is that there is no asymptotic bias in the
limiting null distribution of $\tilde{T}_n$. Another notable feature of
$\tilde{T}_n$ is that the asymptotic variance is also reduced since
$\int K^2(u)du\leq \int[2K(u)-K*K(u)]^2du$. For a formal proof of this inequality, see, e.g, Dette and von Lieres und Wilkau (2001). We will explore this point further in the power performance studies. As for the situation of homoscedastic errors, we have $V_1=|\Omega|{\int K^2(u)du}/{2}.$ Here $|\Omega|$ is also the Lebesgue's measure of the support $Z$. The above theorem shows that under (and only under) conditional homoskedasticity, $V_1$ does not depend on nuisance parameters and nuisance functions.
In this case, the bias-correction version, like the NGLR statistic
$\overline T_n$ in (\ref{fanstatis}), also enjoys the Wilks phenomenon. This offers a great
convenience in implementing the bias-correction version. As the original NGLR test, we need to estimate $V_1$ to
obtain a standardized version of $\tilde{T}_n$. To this end, notice
that
$$\hat V_1=\frac{h\sum_{i=1}^n\sum_{j\neq i}^n \tilde{w}_{ij}^2(\hat B(\hat q))\hat e^2_i\hat e^2_j}
{2\left[{n^{-1}\sum_{i=1}^n \big|\hat e_i[y_i-\hat {\tilde m}(\hat B(\hat q)^\top x_i)]\big|}\right]^2}\Rightarrow V_1,$$ here
$\hat e_i=y_i-g(\hat\beta^\top x_i,\hat\theta)$ and $\tilde{w}_{ij}(\hat B(\hat q))$ has been defined in (\ref{wanwanw}).

We now standardize $\tilde{T}_n$ in (\ref{nobias_statistic}) to get a scale-invariant statistic.
According to Theorem~\ref{theo_nullnobias}, the standardized version of
$\tilde{T}_n$ is
\begin{eqnarray}\label{tildeSn}
R_n:=\frac{\sqrt{h}\tilde{T}_n}{\sqrt{\hat V_1}}=\frac{\sum_{i=1}^n |\hat e_i|\big\{|\hat e_i|-|y_i-\hat {\tilde m}(\hat B(\hat q)^\top x_i)|\big\}}{\sqrt{2\sum_{i=1}^n\sum_{j\neq i}^n \tilde{w}_{ij}^2(\hat B(\hat q))\hat e^2_i\hat e^2_j}}.
\end{eqnarray}
By the consistency of $\hat V_1$, an application of the Slutsky's
Theorem yields the following corollary.
\begin{corollary}\label{coro2}
Under the conditions in Theorem~\ref{theo_nullnobias} and $H_0$, we  have
$$
R_n:=\frac{\sqrt{h}\tilde{T}_n}{\sqrt{\hat V_1}} \Rightarrow N(0,1),
$$
where $N(0,1)$ is the standard normal distribution.
\end{corollary}

From this corollary, we can calculate  $p$-values easily by using
its limiting null distribution.

\subsection{Power study}
Now we are in the position to study the power performance of the test under a sequence of
 local alternative models with the following forms
\begin{equation}\label{loca_alter}
  H_{1n}: Y=g(\beta^\top X,\theta)+C_n m(B^\top X)+\varepsilon.
\end{equation}
Here, $E(\varepsilon|X)=0$, $E[m^2(B^\top X)]<\infty$ and $\{C_n\}$
is a constant sequence. Denote
$\tilde\alpha=(\beta,\theta)$ to be the minimizer
\begin{equation*}
 \tilde\alpha=\arg\min_{\alpha} E\big\{g(\beta^\top X,\theta)-m(X)\big\}^2
\end{equation*}
with $m(X)=E(Y|X)$. Under the null hypothesis, $\tilde{\alpha}$ is the true parameter.
Further, for the least square estimate $\hat\theta$, we always have
$\hat\alpha-\tilde{\alpha}=O_p(1/\sqrt{n})$. We state the consistency  of $\hat q$ under the local
alternative hypothesis (\ref{loca_alter}).

\begin{lemma}\label{conven_loca}
Suppose that conditions (C1)-(C7) in  Appendix hold, under the local alternative (\ref{loca_alter}) with $C_n=n^{-1/2}h^{-1/4}$, and under global alternative (\ref{gen-alter1}), either the OPG-based or the MAVE-based estimate $\hat q$ respectively equals 1 or $q$ with a  probability going to one  as $n\rightarrow\infty$.
\end{lemma}

To state the following theorem, we first define the notation as
\begin{equation}\label{mu1}
  \mu_1=\frac{E[m^2(B^\top X)]}{2\int \sigma^2(z)f(z)dz}.
\end{equation}
The asymptotic properties under global and local alternative
hypothesis are stated in the following Theorem:

\begin{theorem}\label{theo2}
Given  conditions (C1)-(C7) in  Appendix, we have the following results.

(i) Under the global alternative of (\ref{gen-alter1}),
$$\sqrt h (T_n-\frac{Q_1}{h})/(nh^{1/2})\Rightarrow C>0,$$
where $C$ is a positive constant.

(ii) Under the local alternative hypotheses in (\ref{loca_alter})
with $C_n=n^{-1/2}h^{-1/4}$, we have
\begin{equation*}
  \sqrt h (T_n-\frac{Q_1}{h})\Rightarrow N(\mu_1,V_0),
\end{equation*} and $S_n \Rightarrow N(\mu_1/\sqrt{V_0},1)$.
\end{theorem}

This theorem indicates under the global alternative
hypothesis, the proposed test is consistent with the asymptotic
power $1$,  the test can detect the local alternatives distinct
from the null at the rate of order $n^{-1/2}h^{-1/4}$ rather than the rate of order $n^{-1/2}h^{-p/4}$ the NGLR can achieves.
We further  present the asymptotic properties of the bias-corrected version $\tilde{T}_n$ in (\ref{nobias_statistic}).

\begin{theorem}\label{theo_alternobias}
Assume that conditions (C1)-(C7) in  Appendix hold,  we have the following conclusions.

(i) Under the global alternative of (\ref{gen-alter1}),
\begin{equation*}
  \sqrt h \tilde{T}_n/(nh^{1/2})\Rightarrow C_1 >0,
\end{equation*}
where $C_1$ is a positive constant.

(ii) Under the local alternative hypotheses in (\ref{loca_alter})
with $C_n=n^{-1/2}h^{-1/4}$, we have
\begin{equation*}
  \sqrt h \tilde{T}_n\Rightarrow N(\mu_1,V_1),
\end{equation*}
and $R_n \Rightarrow N(\mu_1/\sqrt{V_1},1)$, where $\mu_1, V_1$ have been defined in (\ref{mu1}) and (\ref{V1}), respectively.
\end{theorem}
%
%
Denote
\begin{eqnarray*}
O_1&=&\frac{\mu_1}{\sqrt{V_0}}=\frac{E\{m^2(B^\top X)\}}{\sqrt{2\int\sigma^4(x)dx\int[2\mathcal{K}(u)-\mathcal{K}*\mathcal{K}(u)]^2du}},\\
O_2&=&\frac{\mu_1}{\sqrt{V_1}}=\frac{E\{m^2(B^\top X)\}}{\sqrt{2\int\sigma^4(x)dx\int
  \mathcal{K}^2(u)du}}.
\end{eqnarray*}
From the above theorems, it can be also shown that, the
asymptotic powers of $T_n$ and $\tilde{T}_n$ are
$1-\Phi(z_{\alpha}-O_1)$ and $1-\Phi(z_{\alpha}-O_2)$ respectively
for the alternatives, which are distinct from the null ones at rate
$n^{-1/2}h^{-1/4}$. Since $V_1\leq V_0$, it is evident that
$\tilde{T}_n$ is more powerful than $T_n$ in theoretically. Thus, $\tilde{T}_n$ is asymptotically more efficient than
$T_n$ under $H_{1n}$.

\begin{remark}
Both of our preliminary simulation results based on $S_n$ in the formula (\ref{formulaSn}) and bias-correction statistic $R_n$ in (\ref{tildeSn}) show the inflation sizes of the tests. Therefore, a size-adjustment  is adopted:
\begin{equation}\label{adjuSn}
    \tilde{S}_n=\frac{S_n}{1+4n^{-4/5}}~~~\mbox{and}
    ~~~\tilde{R}_n=\frac{R_n}{1+4n^{-4/5}}.
\end{equation}
Note that the size-adjustment value is asymptotically negligible when $n\rightarrow\infty$ and thus $\tilde{S}_n\rightarrow S_n, \tilde{R}_n\rightarrow R_n$. The size-adjustment is selected via intensive simulation with many different values we conduct and this one is worthy of a recommendation. After such an adjustment, the new tests $\tilde{S}_n,\tilde{R}_n$ can much better control type I errors and enhance the powers than those without the size-adjustment.

\end{remark}

\section{Simulation study}\label{sec4}
\renewcommand{\theequation}{4.\arabic{equation}}
\setcounter{equation}{0}
  Denote the adjusted test statistics in the formula (\ref{adjuSn}) based on
OPG and MAVE methods as $\tilde{S}_n^{OPG}$ and $\tilde{S}_n^{MAVE}$, respectively and the corresponding
bias-corrected versions are denoted as $\tilde{R}_n^{OPG}$ and $\tilde{R}_n^{MAVE}$. In this section, three numerical studies are carried out to examine the finite sample performance of the proposed tests. The first study is used to examine and compare the performance among our proposed four tests. The effect of nonlinearity under the null hypothesis on the performance of the tests is also considered here. The objective of the second study is to examine how much improvement our method can make compared with the NGLR test $T_n^{FZZ}$ proposed by Fan et al (2001) and the impact of correlation between the covariate $X$ is also discussed in this study. Since the test $T_n^{GWZ}$ proposed by Guo et al (2015) is also to solve the dimensionality problem in model checking, the last study aims at comparing our tests with $T_n^{GWZ}$.
%

\vspace{0.3cm}
{\it Study~1}: The data are generated from the following models:
\begin{eqnarray*}
   && H_{11}: Y=\beta^\top X+a_1 \exp(-0.1\beta^\top X)+\varepsilon ~\mbox{and}~\beta=(1,\ldots,1,0,0)^\top/\sqrt{p-2},\\
   && H_{12}: Y=\beta^\top X+1.25 a_1\times 2^{-\beta^\top X}+\varepsilon ~\mbox{and}~\beta=(\underbrace{0,\ldots,0}_{p/2},\underbrace{1,\ldots,1}_{p/2})^\top/\sqrt{p/2},\\
   && H_{13}: Y=\beta^\top X+a_2 \cos(0.6\pi\beta^\top X)+\varepsilon~\mbox{and}~\beta=(1,\ldots,1)^\top/\sqrt{p},\\
   && H_{14}: Y=1.5\exp(0.5\beta^\top X)+a_2\cos(0.6\pi\beta^\top X)+\varepsilon~\mbox{and}~\beta=(1,\ldots,1)^\top/\sqrt{p},
\end{eqnarray*}
where  $p=8$. The covariate
$X=(X_1,X_2,\ldots,X_p)^\top$ are i.i.d. and generated from a multivariate normal distribution $N(0,I_{p})$ where $I_p$ is a $p\times p$ identity matrix. The residual $\varepsilon\sim N(0,1)$ and $\varepsilon\sim t(5)$ are considered. Here, we set $a_1=0,0.1,\ldots,0.5$ and $a_2=0,0.2,\ldots,1.0$. In
these two models, $a_i=0,\,\,\,i=1,2$ corresponds to the null hypothesis and
$a_i\neq 0,\,\,\,i=1,2$ to the alternative hypotheses. Under the alternatives, the last two models are high-frequent and the first two are not. We  examine whether our test can be powerful for these two types of models. Throughout these simulations, unless otherwise specified, the kernel function is taken to be $\mathcal{K}(u)=15/16(1-u^2)^2$ if $|u|\leq 1$ and $0$, otherwise. The bandwidth is
selected as $h=1.5n^{-1/(4+\hat q)}$. The sample sizes $n=100$ and $200$ are considered and the significance level is set to be $\alpha=0.05$. Every simulation result is the average of 2000
replications. Empirical sizes (type I errors) and simulated powers of our test
against the alternatives $H_{1i},\,\,i=1,2,3,4$ are tabulated in Table~\ref{study1_1} and Table~\ref{study1_2} to compare the performance of the four test statistics
$\tilde{S}_n^{OPG},\tilde{R}_n^{OPG}$ and $\tilde{S}_n^{MAVE},\tilde{R}_n^{MAVE}$.

Based on Table~\ref{study1_1}, we can obtain the following
observations. First and the most important, for every combination of the random error $\varepsilon$ and the sample sizes we conduct, both of the size-adjustment bias-correction versions
$\tilde{R}_n^{OPG}$ and $\tilde{R}_n^{MAVE}$ can very well control empirical sizes that are very close to the pre-specified significance level $0.05$. However, the test statistics $\tilde{S}_n^{OPG}$ and $\tilde{S}_n^{MAVE}$ without bias-correction can not always make excellent performance although size-adjustment is done. It is worth mentioning that the empirical sizes of them present too large or too small which can not be adjusted further. Second, with increasing of $a$,  the more deviation of the alternative hypothesis is from the null hypothesis, the higher  the simulated powers are. Also, it is reasonable that the empirical powers of these tests are higher with larger sample sizes. Compared the bias-correction version $\tilde{R}_n^{OPG}$ (or $\tilde{R}_n^{MAVE}$ ) and no bias-correction version  $\tilde{S}_n^{OPG}$ (or $\tilde{S}_n^{MAVE}$), we can see that in most cases, the bias-correction version $\tilde{R}_n^{OPG}$ (or $\tilde{R}_n^{MAVE}$) owns more powerful powers but they are still comparable. As for OPG-based test $\tilde{S}_n^{OPG}$ (or $\tilde{R}_n^{OPG}$) and MAVE-based test $\tilde{S}_n^{MAVE}$ (or $\tilde{R}_n^{MAVE}$),  in most situations, MAVE-based test generally has slightly higher empirical powers than OPG-based test. However, the differences can be negligible, that is, the simulated powers of OPG-based test and MAVE-based test are also comparable. Third, under the same alternative hypothesis, the distribution of random error makes no significant influence and their's empirical sizes and powers are all acceptable, which suggest that our tests are robust. The similar conclusions can be made based on Table~\ref{study1_2} and thus we omit them.

In summary, the above conclusions indicate that the bias-correction is necessary. In the following simulation, we only report the results of OPG-based bias-correction  version $\tilde{R}_n^{OPG}$ for its simplification, light computational burden and comparable performance with $\tilde{R}_n^{MAVE}$.

\begin{center}
Table~\ref{study1_1} and Table~\ref{study1_2} about here
\end{center}

%

\vspace{0.3cm}
Study~2:
Consider the following regression models
\begin{eqnarray*}
  &&H_{21}: Y=\beta_1^\top X+a_1 (\beta_2^\top X)^2+\varepsilon\\
  &&H_{22}: Y=\beta_1^\top X+a_2 |\beta_2^\top X|^{1/2}+\varepsilon,
\end{eqnarray*}
where $p=4$ and $\beta_1=(1,1,\ldots,1)^\top/\sqrt{p}$,
$\beta_2=(\underbrace{0,\ldots,0}_{p/2},1,\ldots,1)^\top/\sqrt{p/2}$, thus when $a_i\neq 0,\,\,i=1,2$, we have $q=2$ and $B=(\beta_1,\beta_2)$. The covariate $X=(X_1,\ldots,X_p)^\top$ are generated from multivariate normal distribution $N(0,\Sigma_k),\,\,k=1,2$ with $\Sigma_1=I_p$ and $\Sigma_{2}=\{0.2^{|i-j|}\}_{p\times p}$. The random error is $\varepsilon\sim N(0,2.56^2)$ and $t(5)$, which is used to examine the effect of the heavy tailed error on the performance of
our proposed test statistics. Denote $a_1=0,0.2,\ldots,1.0$ in $H_{21}$ and $a_2=0,0.3,\ldots,1.5$.
Empirical sizes and powers for $H_{21}, H_{22}$ with $p=4$ and $n=100,200$ are displayed
in Table~\ref{study2}  to compare the results of $\tilde{R}_n^{OPG}$ and $T_n^{FZZ}$. For the NGLR test, just as Hong and Lee (2013) mentioned, the asymptotic normal approximation might
not perform well in finite sample cases due to its slow converge rate to its limiting null distribution. Thus for $T_n^{FZZ}$, except for the asymptotic method $T_{n,A}^{FZZ}$, we also apply the conditional bootstrap procedure $T_{n,B}^{FZZ}$ developed by Hong and Lee (2013)
to determine critical values.

The following findings can be obtained from Table~\ref{study2}.
First, we can see that the empirical sizes of both $\tilde{R}_n^{OPG}$ and
bootstrapped version of $T^{FZZ}_{n,B}$ can be under control and they are robust to various
covariance matrix of $X$ but the asymptotic version   $T^{FZZ}_{n,A}$ can not control sizes very well. Second, It is reasonable that the simulated powers for both the tests become higher with increasing of $a$ and for the same combination of $X$, both the tests are more powerful with larger sample sizes. Third, the proposed test $\tilde{R}_n^{OPG}$ possesses higher powers than $T_n^{FZZ}$ has. $T^{FZZ}_{n,A}$ is failure and it is hard to detect the alternatives when the dimension of $X$ is large. It indicates that the model-adaptive enhancement of NGLR ($\tilde{R}_n^{OPG}$) is not significantly affected by the dimension of the covariate $X$, but the classic  NGLR test inevitably and significantly suffers curse of dimensionality even we use the bootstrap approximation to help.

\begin{center}
Table~\ref{study2} about here
\end{center}

\begin{figure}[htbp]
\centering
\includegraphics[width=14cm,height=12cm]{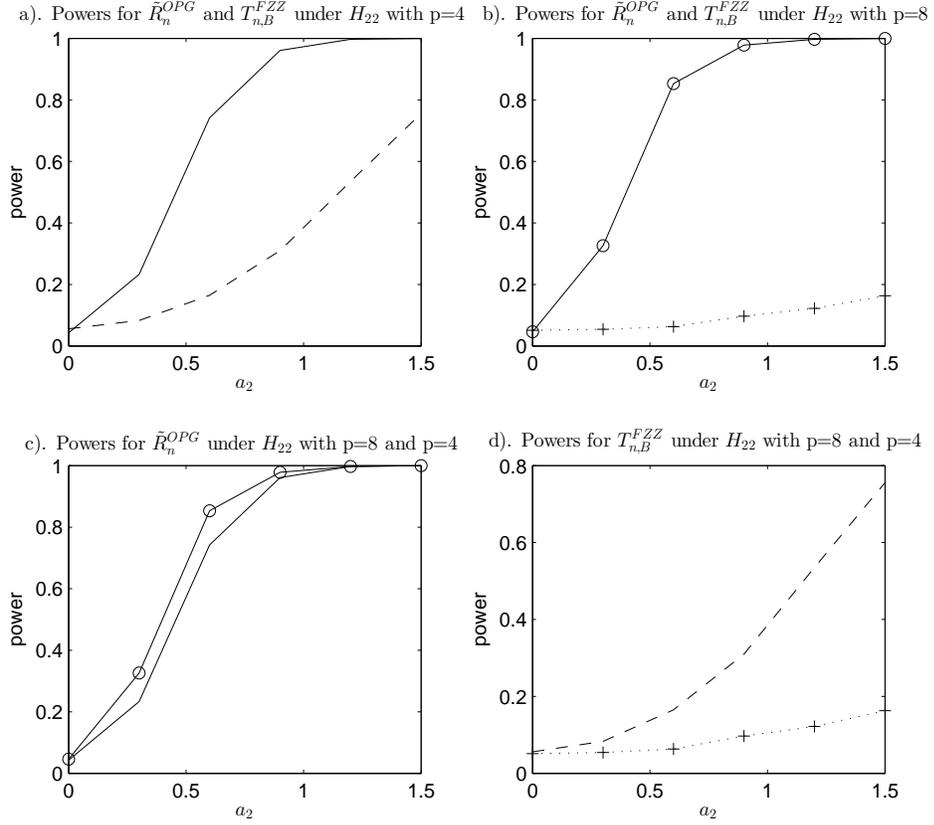}
 \caption{\small
 Empirical sizes and powers of $\tilde{R}_n^{OPG}$ and $T_{n,B}^{FZZ}$ for $H_0$ v.s. $H_{22}$ at significance level $\alpha=0.05$ with $n=100$, $X\sim N(0,I_p)$, $\varepsilon\sim t(5)$. In four plots, the solid line and the dash line are for $\tilde{R}_n^{OPG}$ and $T_{n,B}^{FZZ}$ with $p=4$, respectively. The solid line marked with ``o'' and the dash line marked with ``$+$''is for $\tilde{R}_n^{OPG}$ and $T_{n,B}^{FZZ}$ with $p=8$, respectively.
  }
  \label{p48study2}
\end{figure}

Figure~\ref{p48study2} reports the  power comparisons when  $p=4$ and $p=8$ for the alternative model $H_{12}$. Here, $n=100$, $X\sim N(0,I_p)$, $\varepsilon\sim t(5)$ is considered. Plots a) and b) suggest that in the  $p=4$ and $p=8$ case,  $\tilde{R}_n^{OPG}$ has uniformly better power performance than $T_{n,B}^{FZZ}$. From plots c) and d), we can see that the change of dimensionality has no significant impact on $\tilde{R}_n^{OPG}$ since our test can behave like a local smoothing test as if $X$ were one-dimensional. However,   $T_{n,B}^{FZZ}$ with $p=4$ can be much more powerful than it with $p=8$. 
\begin{center}
Figure~\ref{p48study2} about here
\end{center}

\begin{figure}[htbp]
\centering
\includegraphics[width=14cm,height=12cm]{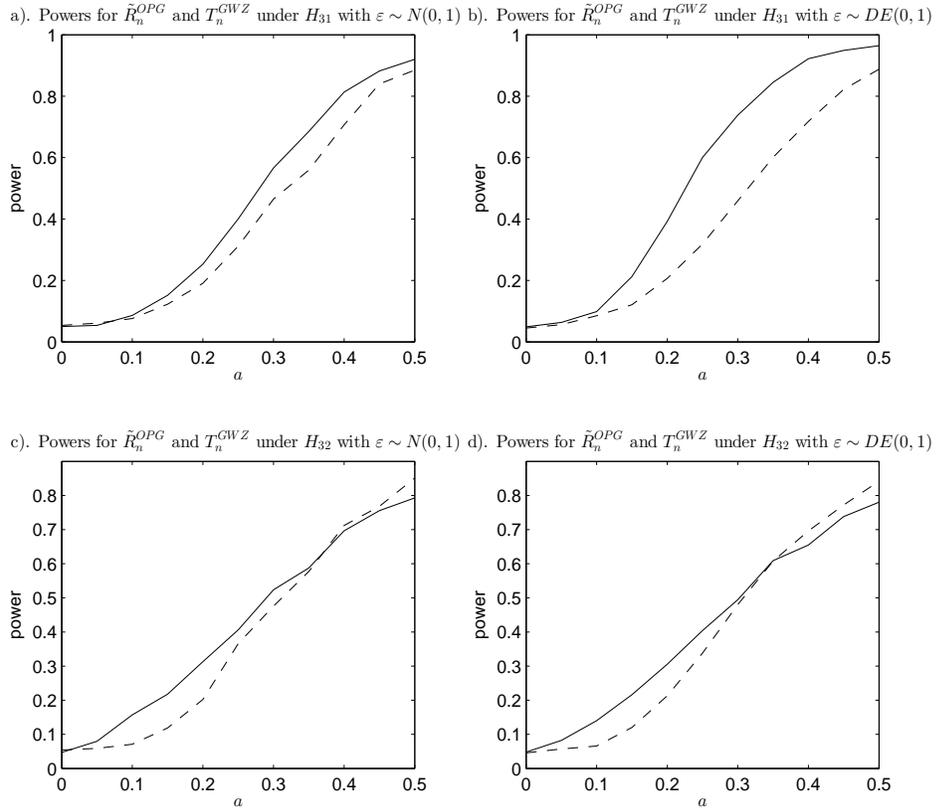}
 \caption{\small
 Empirical sizes and powers of $\tilde{R}_n^{OPG}$ and $T_n^{GWZ}$ for $H_0$ v.s. $H_{31}$ and $H_{32}$ at significance level $\alpha=0.05$ with $n=100$, $p=8$. In four plots, the solid line and the dash line are for $\tilde{R}_n^{OPG}$ and $T_n^{GWZ}$, respectively.
  }
  \label{study3}
\end{figure}

\vspace{0.3cm}
Study~3:
We generate the simulation data from the following regression models
\begin{eqnarray*}
    &&H_{31}: Y=\beta_1^\top X+a_1(\beta_1^\top X)^2+\varepsilon,\\
    &&H_{32}: Y=\beta_1^\top X+a_2(\beta_2^\top X)^3+\varepsilon.
\end{eqnarray*}
For both cases, $p=8$, $\beta_1=(\underbrace{0,\ldots,0}_{p/2},1,\ldots,1)^\top/\sqrt{p/2}$, $\beta_2=(1,1,\ldots,1)^\top/\sqrt{p}$. Here, $X=(X_1,X_2,\ldots,X_p)^\top$  are  generated from the multivariate normal distribution $N(0,I_p)$ where $I_p$ is $p\times p$ identity matrix. The error term
$\varepsilon$ comes from univariate standard normal distribution $N(0,1)$ or Laplace distribution $Laplace(0,1)$ with probability density function $f(x)=\exp(-|x|)/2$. As for the models under $H_{31}$ and $H_{32}$, $a$ is set to be $a_1=0,0.05,0.1,\ldots,0.5$. The simulated sizes and powers with $n=100$ for all of the combinations are plotted in Figure~\ref{study3}, where the test proposed by Guo et al (2015) is denoted as $T_n^{GWZ}$.

From Figure~\ref{study3}, we can see that both the tests $\tilde{R}_n^{OPG}$ and $T_n^{GWZ}$ have acceptable empirical sizes. Further, under all of the situations we conduct, as for the model $H_{31}$, from plots a) and b), we can see that $\tilde{R}_n^{OPG}$ has uniformly higher powers than $T_n^{GWZ}$, though $T_n^{GWZ}$ also performs well. Under the alternative $H_{32}$, from plots c) and d), we can conclude that $\tilde{R}_n^{OPG}$ makes the comparable performance with $T_n^{GWZ}$.
This study suggests that in the limited simulations we conduct, the proposed test can work  better than $T_n^{GWZ}$. As $T_n^{GWZ}$ is also a model-adaptive test, it deserves a further study to see whether this is because NGLR can be more powerful than the one Zheng's test (1996), which is left to
a future research topic.
\begin{center}
Figure~\ref{study3} about here
\end{center}

\section{Real data analysis}
\renewcommand{\theequation}{5.\arabic{equation}}
\setcounter{equation}{0}

A sample of 82 horse mussels are analysed in this section with our proposed test procedure. The data were part of a large ecological study of mussels (Cook, 1998; Cook and Weisberg, 1999), which was collected in the Marlborough Sounds off the coast of New Zealand. Recently, this dataset was used to conduct transformed sufficient dimension by Wang et al (2014). Five variables are included in this dataset: the muscle mass $Y$, the height $H$, the length $L$, the width $W$ and the mass $S$ of the mussel's shell, where $H$, $L$ and $W$ are in millimetres and $S$ is in grams. Before analysis, all of variables are standardized separately. We are interested in whether the relationship between response $Y$ and covariates $X=(H,L,W,S)^\top$ is linear, if not, which model can be suitable to fit this data?

\begin{figure}[htbp]
\centering
\includegraphics[width=14cm,height=12cm]{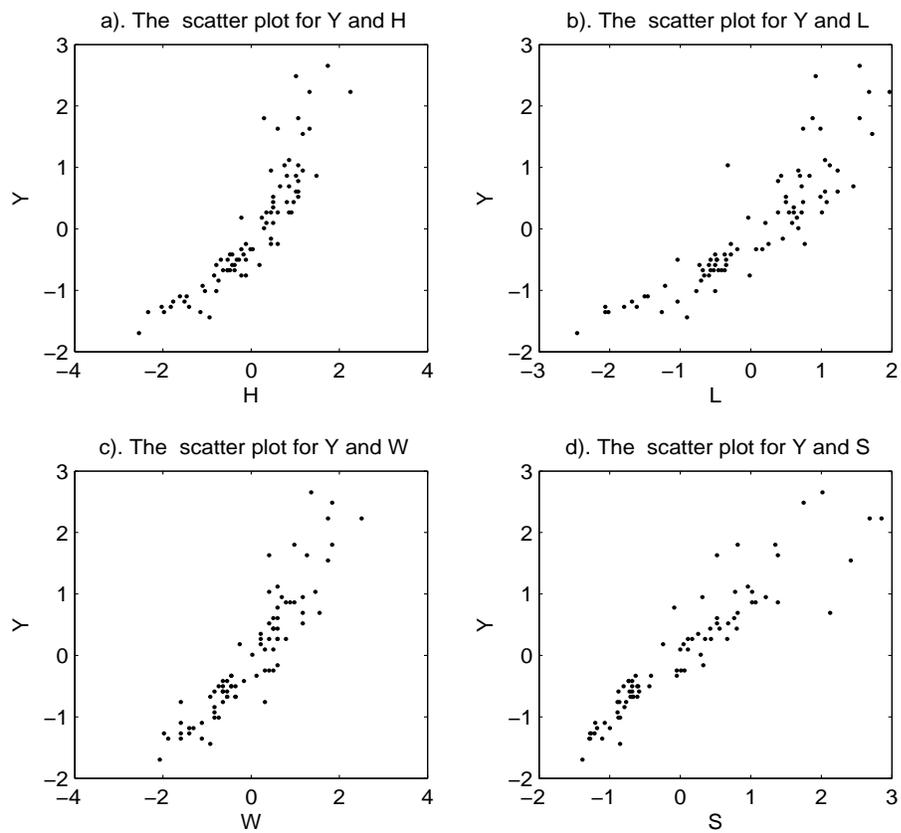}
 \caption{\small
The scatter plots for original response $Y$ and covariate $H,L,W,S$, respectively.
  }
  \label{realorigin}
\end{figure}

Figure~\ref{realorigin} presents the scatter plots for response $Y$ and covariates $H,L,W,S$ and shows that the relationships between $Y$ and $H, L$ are both curved although the variables $W,S$ may be approximately fitted by linear mean functions. We can make a preliminary inference that it is not appropriate to match this dataset with a simple linear regression model. Further, a test is conducted to certify our thought.
The  $p$-values for test statistics $\tilde{R}_n^{OPG}$ and $\tilde{R}_n^{MAVE}$ are both $0.001$, which both suggest that we can reject the null hypothesis in the statistical sense and verify our initial thought.

\begin{center}
Figure~\ref{realorigin} about here
\end{center}

\begin{figure}[htbp]
\centering
\includegraphics[width=14cm,height=12cm]{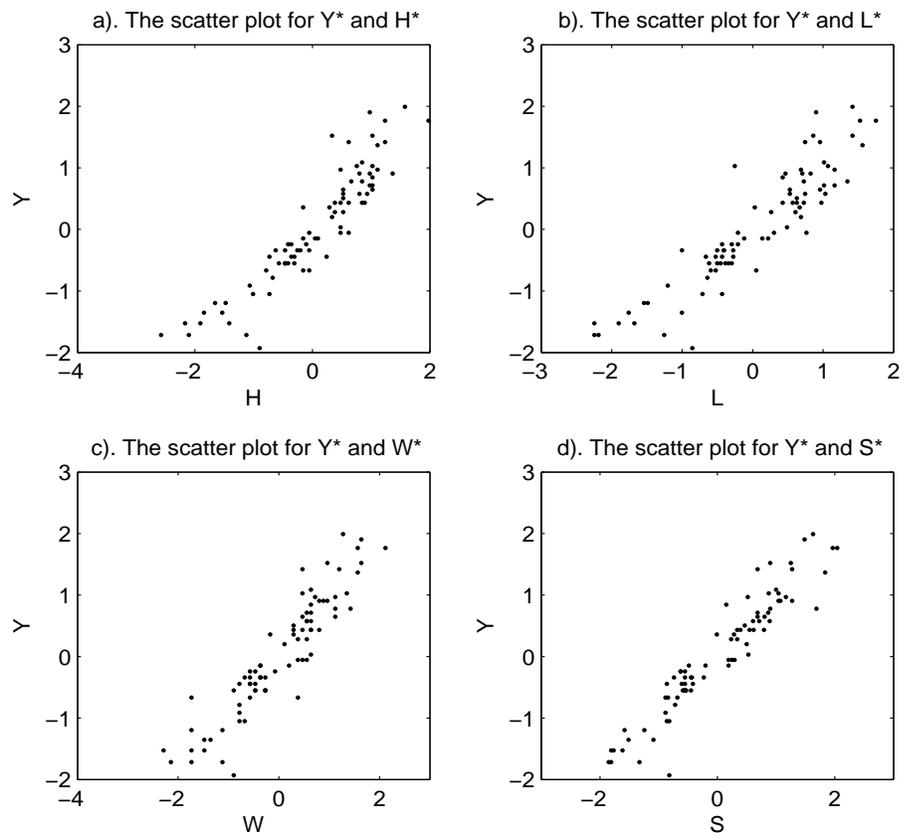}
 \caption{\small
The scatter plots for transformed response $Y^\star$ and covariate $H^\star,L^\star,W^\star,S^\star$, respectively.
  }
  \label{realtrans}
\end{figure}

We further try to implement Yeo-Johnson transformations with the data before analysis. Yeo and Johnson (2000) proposed the following transformation, $\psi(\cdot,\cdot)$: $R\times R\rightarrow R$, where
\begin{eqnarray*}\label{YJtrans}
\psi(\lambda,u)=
\left\{
  \begin{array}{ll}
    \{(u+1)^\lambda-1\}/\lambda & (u\geq 0,\lambda\neq 0), \\
    \log(u+1) & (u\geq 0,\lambda=0),\\
    -\{(-u+1)^{2-\lambda}-1\}/(2-\lambda) & (u< 0,\lambda\neq 2),\\
    -\log\{(-u+1)\} & (u<0,\lambda=2).
  \end{array}
\right.
\end{eqnarray*}
Denote the transformed response as $\tilde{Y}$ and transformed covariates as $(\tilde{H},\tilde{L},\tilde{W},\tilde{S})^\top$. Here, $\lambda=0.3$ is considered. Since the response and covariates are all positive, only $\psi(\lambda,u)=\{(u+1)^\lambda-1\}/\lambda$ is enough to transform original data.
The scatter plots for transformed response $Y^\star$ and covariates $\tilde{X}=(\tilde{H},\tilde{L},\tilde{W},\tilde{S})^\top$ are depicted in Figure~\ref{realtrans}, which all approximately display linear relationship between them. Further, we  intend to apply the linear regression  $\tilde{Y}={\tilde{X}}^\top\hat\beta$ to fit this dataset and $\hat\beta=(0.256,-0.025,0.104,0.634)^\top$ are obtained. Our proposed test is employed to check whether this model is adequate. The test statistics values for $\tilde{R}_n^{OPG}$ and $\tilde{R}_n^{MAVE}$ are $-1.697$ and $-1.742$. The corresponding $p$-values are $0.090$ and $0.082$, which both indicate that the null hypothesis cannot be rejected provided that the significance level $\alpha=0.05$ is considered, that is, the linear regression model is proper to fit the transformed data.

\begin{center}
Figure~\ref{realtrans} about here
\end{center}

\section*{Appendix. Proofs of  theorems}
\renewcommand{\theequation}{A.\arabic{equation}}
\setcounter{equation}{0}

The following conditions are required for proving the theorems in Section~3.
\begin{itemize}
  \item [(C1)] $E|Y|^k<\infty$, $E||X||_2^k<\infty$ for all $k>0$, $E|\varepsilon|<\infty$, $E|g(\beta^\top X,\theta)|<\infty$, $E\{[m(B^\top X)-g(\beta^\top X,\theta)]^2\}<\infty$,$E|m(B^\top X)|<\infty$, $\sup E(X_l^2|B^\top X)<\infty,\,\,l=1,2,\ldots,p$; $E(\varepsilon^2|B^\top X)<\infty$ where $\varepsilon=Y-E(Y|B^\top X)$; $E(X|Y)$ and $E(XX^\top|Y)$ have bounded, continuous third order derivatives.

  \item [(C2)] The density function $f_\beta(z)$ of $\beta^\top X$ on support of $Z$ exists and  has bounded derivatives up to order $r,\,\,r\geq 2$ for all $\beta$: $|\beta-\beta|<\delta$ where $\delta>0$ and satisfies
      \begin{equation*}
        0<\inf f(z)<\sup f(z)<1.
      \end{equation*}

  \item [(C3)] The density function $f(X)$ of $X$ has bounded second derivatives and is abounded away from $0$ in a neighbor around $0$. The density function $f(Y)$ of $Y$ has bounded derivative and is bounded away from $0$ on a compact support. The conditional densities $f_{X|Y}(\cdot)$ of $X$ given $Y$ and $f_{(X_0,X_l)|(Y_0,Y_l)}$ of $(X_0,X_l)$ given $(Y_0,Y_l)$ are bounded for all $l\geq 1$.

  \item [(C4)] The conditional mean $E(Y|\beta^\top X=z)$ has bounded derivatives up to order $r,\,\,r\geq 2$ for all $\beta$: $|\beta-\beta|<\delta$ where $\delta>0$. At the same time, the derivatives of link functions $g(\cdot)$ and $h(\cdot)$ up to order $3$ are bounded.

  \item [(C5)] The kernel function $\mathcal{K}(\cdot)$ is a bounded, derivative and symmetric probability density function which satisfies the Lipschitz condition and all the moments of $\mathcal{K}(\cdot)$ exist. The bandwidth and trimming parameter satisfy $h\propto n^{-1/5}$ and $\epsilon<1/20$ respectively. Besides, $\sqrt{nh}\rightarrow\infty$ and $nh^{1/2 +2r}\rightarrow\infty$.

  \item [(C6)] Under the null hypothesis and local alternative hypothesis, $nh^2\rightarrow\infty$; Under the global alternative hypothesis, $nh^q\rightarrow\infty$.

  \item [(C7)] The matrix $E\{\nabla m(B^\top X)\nabla m(B^\top X)^\top\}$ is positive definite where $\nabla m(\cdot)=m'(\cdot)$ denotes the gradient of the function $m(\cdot)$.
\end{itemize}

\begin{remark}\label{remark_condition}
  Condition (C1), Condition (C3) and Condition (C5) are required for  MAVE. Condition (C2) is necessary for the asymptotic normality of our test statistics. The smoothness requirements on the link function in Condition (C4) can be relaxed to the existence of a bounded second-order derivative at the cost of more complicated technical proofs and the use of smaller bandwidth, which is similar to the conditions in Xia (2006). Conditions (C5) and  (C7) are assumed for OPG. In Condition (C6), $nh^2\rightarrow\infty$ is a common assumption in nonparametric estimation.
\end{remark}

We first give the proof of Lemma~\ref{conven_loca} in Section 3.

\vspace{0.3cm} \noindent \textit{Proof of Lemma~\ref{conven_loca}.}
We only prove this lemma for the OPG based estimate. The result for
MAVE based estimate has been proven by Guo, Wang and Zhu (2015). Let
$G(B^\top X)=g(\tilde{\beta}^\top X,\tilde{\theta})+C_n m(B^\top
X)$. Thus
$$\nabla G(B^\top X)=g'(\tilde{\beta}^\top X,\tilde{\theta})\tilde{\beta}
+C_n B m'(B^\top X).$$

It then follows that:
\begin{eqnarray*}
E\{\nabla G(B^\top X)\nabla
G(B^\top X)^\top\}&=&E\{g'(\tilde{\beta}^\top X,\tilde{\theta})^2\}
\tilde{\beta}\tilde{\beta}^\top+2C_n\tilde{\beta}E\{g'(\tilde{\beta}^\top
X,\tilde{\theta})m'(B^\top X)^\top\}B^\top\\
&&+C^2_nBE\{m'(B^\top X)m'(B^\top X)^\top\}B^\top.
\end{eqnarray*}
Denote $\tilde{\Sigma}=n^{-1}\sum_{j=1}^n b_jb^\top_j$. We  have:
\begin{eqnarray*}
\hat\Sigma-E\{g'(\tilde{\beta}^\top X,\tilde{\theta})^2\}
\tilde{\beta}\tilde{\beta}^\top&=&\hat\Sigma-\tilde{\Sigma}+\tilde{\Sigma}-E\{\nabla
G(B^\top X)\nabla G(B^\top X)^\top\}\\
&&+E\{\nabla G(B^\top X)\nabla G(B^\top
X)^\top\}-E\{g'(\tilde{\beta}^\top X,\tilde{\theta})^2\}
\tilde{\beta}\tilde{\beta}^\top\\
&=&\frac{2}{n}\sum_{j=1}^n (\hat
b_j-b_j)b^\top_j+O_p(\frac{1}{\sqrt{n}}+C_n)=O_p(\frac{1}{\sqrt{n}}+C_n).
\end{eqnarray*}
Thus, we can get
$\hat\lambda_i-\lambda_i=O_p(\frac{1}{\sqrt{n}}+C_n)$. Note that
$\lambda_1>0$ and for any $l > 1$, we have $\lambda_l = 0$.
Consequently, under the condition that $C_n=o(c)$ and $c=o(1)$,
\begin{eqnarray*}
\frac{\hat\lambda_2+c}{\hat\lambda_1+c}-\frac{\hat\lambda_{l+1}+c}{\hat\lambda_l+c}&=&
\frac{\lambda_2+c+O_p(C_n)}{\lambda_1+c+O_p(C_n)}-\frac{\lambda_{l+1}+c+O_p(C_n)}{\lambda_l+c+O_p(C_n)}\\
&=&\frac{c+O_p(C_n)}{\lambda_1+c+O_p(C_n)}-\frac{c+O_p(C_n)}{c+O_p(C_n)}\Rightarrow
-1.
\end{eqnarray*}
Thus under the local alternative (\ref{loca_alter}), Lemma 1 holds.

Further consider the case that under the global alternative (\ref{gen-alter1}). We have
\begin{eqnarray*}
  &&\hat\Sigma-E\{\nabla m(B^\top X)\nabla m(B^\top X)^\top\}\\
  &=&\hat\Sigma-\tilde\Sigma+\tilde\Sigma-E\{\nabla m(B^\top X)\nabla m(B^\top X)^\top\}\\
  &=&\frac{2}{n}\sum_{j=1}^n (\hat
b_j-b_j)b^\top_j+O_p(\frac{1}{\sqrt{n}})=O_p(\frac{1}{\sqrt{n}}).
\end{eqnarray*}
Therefore, $\hat\lambda_i-\lambda_i=O_p(1/\sqrt n)$. Note that when $l\leq q$, $\lambda_l>0$ and
for any $l>q$, we have $\lambda_l=0$. Noticing that $c=1/\sqrt{nh}$ is recommended, thus $1/\sqrt n=o(c)$ and $c=o(1)$. For $l>q$,
\begin{eqnarray*}
  \frac{\hat\lambda_{q+1}+c}{\hat\lambda_{q}+c}-\frac{\hat\lambda_{l+1}+c}{\hat\lambda_{l}+c}
  &=&\frac{\lambda_{q+1}+c+O_p(\frac{1}{\sqrt n})}{\lambda_{q}+c+O_p(\frac{1}{\sqrt n})}-\frac{\lambda_{l+1}+c+O_p(\frac{1}{\sqrt n})}{\lambda_{l}+c+O_p(\frac{1}{\sqrt n})}
  \\
  &=&\frac{c+O_p(\frac{1}{\sqrt n})}{\lambda_{q}+c+O_p(\frac{1}{\sqrt n})}-\frac{c+O_p(\frac{1}{\sqrt n})}{c+O_p(\frac{1}{\sqrt n})}\Rightarrow -1.
\end{eqnarray*}
when $1\leq l<q$,
\begin{eqnarray*}
\frac{\hat\lambda_{q+1}+c}{\hat\lambda_{q}+c}-\frac{\hat\lambda_{l+1}+c}{\hat\lambda_{l}+c}
&=&\frac{\lambda_{q+1}+c+O_p(\frac{1}{\sqrt n})}{\lambda_{q}+c+O_p(\frac{1}{\sqrt n})}-\frac{\lambda_{l+1}+c+O_p(\frac{1}{\sqrt n})}{\lambda_{l}+c+O_p(\frac{1}{\sqrt n})}
  \\
  &=&\frac{c+O_p(\frac{1}{\sqrt n})}{\lambda_{q}+c+O_p(\frac{1}{\sqrt n})}-\frac{\lambda_{l+1}+c+O_p(\frac{1}{\sqrt n})}{\lambda_{l}+c+O_p(\frac{1}{\sqrt n})}
  \Rightarrow -\frac{\lambda_{l+1}}{\lambda_{l}}<0.
\end{eqnarray*}
Therefore, we can conclude that under the global alternative (\ref{gen-alter1}), $\hat q= q$ holds with a probability going to one. The proof is finished. $\Box$

\vspace{0.3cm}
The following lemmas are used to prove the theorems  in
Section~3.
\begin{lemma}\label{lema1}
Under Conditions (C1)-(C7), we have:
\begin{equation*}
  \frac{1}{n}RSS_1=\int \sigma^2(z)f(z)dz+o_p(1),
\end{equation*}
where $Z=B^\top X$.
\end{lemma}

\vspace{0.3cm}
\noindent \textit{Proof of Lemma~\ref{lema1}.}
 The term $RSS_1/n$ can be written as
\begin{eqnarray}
   \frac{1}{n}RSS_1&=&\frac{1}{n}\sum_{i=1}^n\Big\{\varepsilon_i+m(B^\top x_i)-\hat m (\hat B^\top x_i)\Big\}^2\nonumber\\
   &=&\frac{1}{n}\sum_{i=1}^n\varepsilon_i^2+\frac{2}{n}\sum_{i=1}^n\big\{m(B^\top x_i)-\hat m (\hat B^\top x_i)\big\}\varepsilon_i+\frac{1}{n}\sum_{i=1}^n\big\{m(B^\top x_i)-\hat m (\hat B^\top x_i)\big\}^2\nonumber\\
   &=:&A_1+A_2+A_3.\label{RSS_1}
\end{eqnarray}
Let $Z=B^\top X$. The term $A_1$ has the following representation:
\begin{equation}\label{A_1}
  A_1=E[E(\varepsilon^2|Z)]+o_p(1)=\int \sigma^2(z)f(z)dz+o_p(1).
\end{equation}
For the term $A_3$, it can be easily derived that
\begin{eqnarray}
  A_3&\leq&\frac{1}{n}\sum_{i=1}^n \big\{m(B^\top x_i)- m (\hat B^\top x_i)\big\}^2
  +\frac{1}{n}\sum_{i=1}^n \big\{m(\hat B^\top x_i)- \hat m (\hat B^\top x_i)\big\}^2\nonumber\\
  &=:&A_{31}+A_{32}.\label{A_3}
\end{eqnarray}
The result of  H$\ddot{a}$rdle et al. (2004) shows that $\|B-\hat B\|=O_p(1/\sqrt n)$. Note that $\sum_{i=1}^n\|x_i\|^2=O_p(n)$. Then
\begin{equation}\label{A_31}
  |A_{31}|\leq m'^2(\tilde z)\|B-\hat B\|^2\cdot\frac{1}{n}\sum_{i=1}^n\|x_i\|^2+o_p(\frac{1}{n})=O_p(\frac{1}{n}),
\end{equation}
where $\tilde z$ lies between $B^\top x_i$ and $\hat B^\top x_i$. By Theorem~4.8 in H$\ddot{a}$rdle et al. (2004), we can derive
\begin{eqnarray}
  |A_{32}|&\leq& c\cdot \frac{1}{n}\sum_{i=1}^n\Big\{\big[\hat m(\hat B^\top x_i)-E \hat m(\hat B^\top x_i)\big]^2+\big[E\hat m(\hat B^\top x_i)-m(\hat B^\top x_i)\big]^2\Big\}\cdot \hat f(\hat B^\top x_i)\nonumber\\
  &=&O_p(\frac{1}{nh}+h^{2r}).\label{A_32}
\end{eqnarray}
Combining (\ref{A_3}) (\ref{A_31}) and (\ref{A_32}) yield that
\begin{equation}\label{resu_A3}
  A_3=O_p(\frac{1}{n})+O_p(\frac{1}{nh}+h^{2r}).
\end{equation}
Similar to $A_3$, the term  $A_2$ can be bounded by
\begin{eqnarray}
  |A_2|&\leq& \frac{2}{n}\sum_{i=1}^n\Big\{|\hat m(\hat B^\top x_i)-E \hat m(\hat B^\top x_i)|+|E\hat m(\hat B^\top x_i)-m(\hat B^\top x_i)|\Big\}\cdot|\varepsilon_i|\nonumber\\
  &=&O_p(\frac{1}{\sqrt{nh}}+h^r).\label{A_2}
\end{eqnarray}
Taking the formulae (\ref{A_1}), (\ref{resu_A3}) and (\ref{A_2})
into (\ref{RSS_1}), as $nh\rightarrow\infty$, $h\rightarrow 0$, we
have $RSS_1/n$ converge to a constant in probability:
\begin{equation*}
  \frac{1}{n}RSS_1=\int \sigma^2(z)f(z)dz+o_p(1).
\end{equation*}
The proof of Lemma~\ref{lema1} is finished. $\Box$

\vspace{0.5cm}
\begin{lemma}\label{lema2}
Denote $\alpha=(\beta,\theta)$ and let $\tilde\alpha$ be a parameter
value which minimizes $E\big\{g(\beta^\top X,\theta)-m(X)\big\}^2$.
Suppose Conditions  (C1)-(C7) hold. Then
\begin{equation*}
  \hat\alpha-\tilde\alpha=O_p(\frac{1}{\sqrt n}).
\end{equation*}
\end{lemma}

\vspace{0.3cm}
\noindent \textit{Proof of Lemma~\ref{lema2}.}
Note that $\hat\alpha$ can be estimated through OLS, that is,
\begin{equation*}
  \hat\alpha=\min_{(\beta,\theta)} \frac{1}{n}\sum_{i=1}^n \big\{y_i-g(\beta^\top x_i,\theta)\big\}^2.
\end{equation*}
Further, denote $g(\alpha):=g(\beta^\top x_i,\theta)$, we can get
\begin{eqnarray*}
  0&=&\frac{1}{n}\sum_{i=1}^n \frac{\partial g(\alpha)}{\partial\alpha}\Big|_{\alpha=\hat\alpha}(y_i-g(\hat\alpha))\\
  &=&\frac{1}{n}\sum_{i=1}^n  \frac{\partial g(\alpha)}{\partial\alpha}\Big|_{\alpha=\tilde\alpha}(y_i-g(\tilde\alpha))+\Big\{
  \frac{1}{n}\sum_{i=1}^n(y_i-g(\alpha_1))\frac{\partial^2 g(\alpha)}{\partial\alpha\partial\alpha^\top}\Big|_{\alpha=\alpha_1}\\
  &&-\frac{1}{n}\sum_{i=1}^n\frac{\partial g(\alpha)}{\partial\alpha}\Big|_{\alpha=\alpha_1}\frac{\partial g(\alpha)}{\partial\alpha^\top}\Big|_{\alpha=\alpha_1} \Big\}(\hat\alpha-\tilde\alpha),
\end{eqnarray*}
where $\alpha_1$ lies between $\hat\alpha$ and $\tilde{\alpha}$.
According to the above formula, we can obtain that:
\begin{eqnarray}\label{cha_alpha}
  \hat\alpha-\tilde\alpha&=&\Big\{
  \frac{1}{n}\sum_{i=1}^n\frac{\partial g(\alpha)}{\partial\alpha}\Big|_{\alpha=\alpha_1}\frac{\partial g(\alpha)}{\partial\alpha^\top}\Big|_{\alpha=\alpha_1}
  - \frac{1}{n}\sum_{i=1}^n(y_i-g(\alpha_1))\frac{\partial^2 g(\alpha)}{\partial\alpha\partial\alpha^\top}\Big|_{\alpha=\alpha_1}\Big\}^{-1}\nonumber\\
  &&\times \frac{1}{n}\sum_{i=1}^n  \frac{\partial g(\alpha)}{\partial\alpha}\Big|_{\alpha=\tilde\alpha}(y_i-g(\tilde\alpha))\nonumber\\
  &=:&W_1^{-1}W_2.
\end{eqnarray}
It is not difficult to verify that
\begin{equation}\label{W1}
  W_1=E[g'(\tilde\alpha)g'(\tilde\alpha)^\top]+o_p(1),
\end{equation}
where $g'(\cdot)={\partial g(\alpha)}/{\partial\alpha}$ denotes the
gradient of the function $g(\cdot)$. Afterwards, $W_2$ can be
rewritten as
\begin{eqnarray*}
  W_2&=&\frac{1}{n}\sum_{i=1}^n  \frac{\partial g(\alpha)}{\partial\alpha}\Big|_{\alpha=\tilde\alpha}(m(x_i)-g(\tilde\alpha))+
  \frac{1}{n}\sum_{i=1}^n  \frac{\partial g(\alpha)}{\partial\alpha}\Big|_{\alpha=\tilde\alpha}\varepsilon_i.
\end{eqnarray*}
Recall the definition of $\tilde\alpha$, we can derive that
$E(W_2)=0$ and $Var(W_2)=O_p(1/n)$. Further, we get
\begin{equation}\label{W2}
  W_2=O_p(\frac{1}{\sqrt n}).
\end{equation}
Combining (\ref{cha_alpha}), (\ref{W1}) and (\ref{W2}), we can
easily conclude that
\begin{equation*}
  \hat\alpha-\tilde\alpha=O_p(\frac{1}{\sqrt n}).
\end{equation*}
The proof of Lemma~\ref{lema2} is completed. $\Box$

\vspace{0.5cm}
\begin{lemma}\label{lema4}
Suppose Conditions  (C1)-(C7) hold and under either $H_0$  or $H_{1n}$ in (\ref{loca_alter}) with $C_n\rightarrow 0$ as $n\rightarrow\infty$, we have
\begin{equation*}
  \frac{1}{n}\widetilde{\mbox{RSS}}_1=\int \sigma^2(z)f(z)dz+o_p(1),
\end{equation*}
where $Z=B^\top X.$
\end{lemma}

\vspace{0.3cm}
\noindent \textit{Proof of Lemma~\ref{lema4}.}
Let $\Delta({x}_i)= m(B^\top x_i)-g(\tilde{\beta}^\top
x_i,\tilde\theta)$, where $\tilde\alpha=(\tilde\beta,\tilde\theta)^\top$ is defined in Lemma~\ref{lema2}, hence, $y_i=\varepsilon_i+m(x_i)=\varepsilon_i+\Delta({x}_i)+g(\tilde{\beta}^\top
{x}_i,\tilde\theta)$.  consider the following decomposition:
\begin{eqnarray}
  &&[y_i-\hat {\tilde m}(\hat B(\hat q)^\top x_i)][y_i-g(\hat\beta^\top x_i,\hat\theta)]\nonumber\\
  &=&\big[\varepsilon_i+m(B^\top x_i)-\hat {\tilde m}(\hat B(\hat q)^\top x_i)\big]\big[\varepsilon_i+\Delta({x}_i)+g(\tilde{\beta}^\top
{x}_i,\tilde\theta)-g(\hat\beta^\top x_i,\hat\theta)\big]\nonumber\\
&=& \varepsilon_i^2+ \varepsilon_i \Delta({x}_i)+
 \varepsilon_i[g(\tilde{\beta}^\top{x}_i,\tilde\theta)-g(\hat\beta^\top x_i,\hat\theta)]
+ \varepsilon_i[m(B^\top x_i)-\hat {\tilde m}(\hat B(\hat q)^\top x_i)]\nonumber\\
&&+ \Delta({x}_i)[m(B^\top x_i)-\hat {\tilde m}(\hat B(\hat q)^\top x_i)]\nonumber\\
&&+[m(B^\top x_i)-\hat {\tilde m}(\hat B(\hat q)^\top x_i)][g(\tilde{\beta}^\top
{x}_i,\tilde\theta)-g(\hat\beta^\top x_i,\hat\theta)]\nonumber\\
&=:&\tilde{A}_1+\tilde{A}_2+\tilde{A}_3+\tilde{A}_4+\tilde{A}_5+\tilde{A}_6.\label{tildeA16}
\end{eqnarray}
From Lemma \ref{lema2}, $(\hat\alpha-\tilde\alpha)=O_p(1/\sqrt{n})$.
Thus, for the term
$\tilde{A}_3$, we have:
\begin{eqnarray}\label{tildeA3}
  \tilde{A}_3=\varepsilon_i g'(\alpha_2)(\tilde{\alpha}-\hat\alpha)
  =O_p(\frac{1}{\sqrt{n}}),
\end{eqnarray}
where $\alpha_2$ lies between $\hat\alpha$ and $\alpha_0$.

As for the term $\tilde{A}_4$, similarly as the proof of the term $A_2$ in (\ref{A_2}), we can obtain
\begin{equation}\label{tildeA4}
  |\tilde{A}_4|\leq O_p(\frac{1}{\sqrt{nh}}+h^r),
\end{equation}
thus, when $n\rightarrow\infty,h\rightarrow 0,nh\rightarrow\infty$, we have $\tilde{A}_4=o_p(1)$. Further, $\tilde{A}_5=o_p(1)$ can be gotten as well.

Combining (\ref{tildeA3}) and (\ref{tildeA4}), we can conclude $\tilde{A}_6=o_p(1)$.
Through the above proof, the formula (\ref{tildeA16}) can be rewritten as
\begin{equation}\label{ymyg}
  [y_i-\hat {\tilde m}(\hat B(\hat q)^\top x_i)][y_i-g(\hat\beta^\top x_i,\hat\theta)]
 = \varepsilon_i^2+ \varepsilon_i \Delta({x}_i)+o_p(1).
\end{equation}

Consider the two cases under the null and alternative hypothesis.\\
A). Under the null hypothesis $H_0$, $\Delta({x}_i)=0$, then for $i=1,\ldots,n$, when $n\rightarrow\infty$, we have $[y_i-\hat {\tilde m}(\hat B(\hat q)^\top x_i)][y_i-g(\hat\beta^\top x_i,\hat\theta)]= \varepsilon_i^2+o_p(1)>0$, thus, for large $n$,
\begin{eqnarray*}
  \frac{1}{n}\widetilde{\mbox{RSS}}_1&=&\frac{1}{n}\sum_{i=1}^n \Big|[y_i-\hat {\tilde m}(\hat B(\hat q)^\top x_i)][y_i-g(\hat\beta^\top x_i,\hat\theta)]\Big|\\
  &=&\frac{1}{n}\sum_{i=1}^n [y_i-\hat {\tilde m}(\hat B(\hat q)^\top x_i)][y_i-g(\hat\beta^\top x_i,\hat\theta)]+o_p(1)\\
  &=&\frac{1}{n}\sum_{i=1}^n \varepsilon_i^2+o_p(1)=\int\sigma^2(z)f(z)dz+o_p(1).
\end{eqnarray*}

B). Under the local alternative $H_{1n}$ in (\ref{loca_alter}), $\Delta({x}_i)=C_n m(B^\top x_i)$,
\begin{equation*}
  \frac{1}{n}\widetilde{\mbox{RSS}}_1=\frac{1}{n}\sum_{i=1}^n |\varepsilon_i^2+C_n \varepsilon_i m(B^\top x_i)|+o_p(1).
\end{equation*}
Since $C_n\rightarrow 0$ when $n\rightarrow\infty$,
it can be easily obtained that $\Delta({x}_i)\rightarrow 0$, further we have
$$\frac{1}{n}\widetilde{\mbox{RSS}}_1=\int\sigma^2(z)f(z)dz+o_p(1).$$
The proof of Lemma~\ref{lema4} is finished. $\Box$

\vspace{0.5cm} \noindent \textit{Proof of Theorem~\ref{theo1}.}
Under the null hypothesis in (\ref{null_model}), we have
$m(x)=g(\beta^\top x,\theta)$. Write the numerator of $T_n$ in (\ref{statistic}) as  $H_n=(RSS_0-RSS_1)/n$.
Then
\begin{eqnarray}
  H_n&=&\frac{1}{n}\sum_{i=1}^n [y_i-g(\hat\beta^\top x_i,\hat\theta)]^2-\frac{1}{n}\sum_{i=1}^n[y_i-\hat m (\hat B^\top x_i)]^2\nonumber\\
  &=&\frac{1}{n}\sum_{i=1}^n [2y_i-g(\hat\beta^\top x_i,\hat\theta)-\hat m (\hat B^\top x_i)][\hat m (\hat B^\top x_i)-g(\hat\beta^\top x_i,\hat\theta)]\nonumber\\
  &=&\frac{1}{n}\sum_{i=1}^n \big[2\varepsilon_i+g(\beta^\top x_i,\theta)-g(\hat\beta^\top x_i,\hat\theta)+m(x_i)-\hat m (\hat B^\top x_i)\big]\nonumber\\
  &&\big[\hat m (\hat B^\top x_i)-m(x_i)+g(\beta^\top x_i,\theta)-g(\hat\beta^\top x_i,\hat\theta)\big]\nonumber\\
  &=&\frac{1}{n}\Big\{2\sum_{i=1}^n \varepsilon_i[\hat m (\hat B^\top x_i)-m(x_i)]+2\sum_{i=1}^n \varepsilon_i[g(\beta^\top x_i,\theta)-g(\hat\beta^\top x_i,\hat\theta)]\nonumber\\
  &&+\sum_{i=1}^n [g(\beta^\top x_i,\theta)-g(\hat\beta^\top x_i,\hat\theta)]^2-\sum_{i=1}^n[m(x_i)-\hat m (\hat B^\top x_i)]^2
  \Big\}\nonumber\\
  &=:&B_1+B_2+B_3-B_4. \label{fenzi}
\end{eqnarray}

As to the term $B_2$, similarly as the proof of $\tilde{A}_3$ in (\ref{tildeA3}), we have
\begin{eqnarray}\label{truB2}
  B_2=\frac{2}{n}\sum_{i=1}^n\varepsilon_i g'(\alpha_2)(\alpha_0-\hat\alpha)=O_p(\frac{1}{n}),
\end{eqnarray}
where $\alpha_2$ lies between $\hat\alpha$ and $\alpha_0$.

For the term $B_3$, we have:
\begin{equation}\label{truB3}
  B_3=(\hat\alpha-\alpha_0)^\top\cdot\frac{1}{n}\sum_{i=1}^n g'(\tilde\alpha_2)^\top g'(\tilde\alpha_3)(\hat\alpha-\alpha_0)=O_p(\frac{1}{n}),
\end{equation}
where $\tilde{\alpha}_3$ is between $\hat\alpha$ and $\alpha_0$.

Consider the term $B_4$. Recalling the formula (\ref{hatm}), we can obtain
\begin{eqnarray}
  B_4&=&\frac{1}{n}\sum_{i=1}^n\Big\{\sum_{j=1}^nw_{ij}(\hat B)(y_j-m(x_i))\Big\}^2\nonumber\\
  &=&\frac{1}{n}\sum_{i=1}^n\Big\{\sum_{j=1}^nw_{ij}(\hat B)[m(x_j)+\varepsilon_j-m(x_i)]\Big\}^2\nonumber\\
  &=&\frac{1}{n}\sum_{i=1}^n\sum_{j=1}^nw_{ij}(\hat B)\varepsilon_j\sum_{k=1}^nw_{ik}(\hat B)\varepsilon_k
  +\frac{2}{n}\sum_{i=1}^n\sum_{j=1}^nw_{ij}(\hat B)[m(x_j)-m(x_i)]\sum_{k=1}^nw_{ik}(\hat B)\varepsilon_k\nonumber\\
  &&+\frac{1}{n}\sum_{i=1}^n\sum_{j=1}^nw_{ij}(\hat B)[m(x_j)-m(x_i)]\sum_{k=1}^nw_{ik}(\hat B)[m(x_k)-m(x_i)]\nonumber\\
  &=:&B_{41}+B_{42}+B_{43}.\label{B4}
\end{eqnarray}
For the term $B_{42}$,
we first calculate the following element: for any function of
$x$, $l(x)$, we have:
\begin{eqnarray}
    l(x_i)-\sum_{j=1}^n w_{ij}(B)l(x_j)&=&\sum_{j=1}^n w_{ij}( B)\{l(x_i)-l(x_j)\}\nonumber\\
    &=&\frac{1}{nhf(z_i)}\sum_{j=1}^n \mathcal{K}\Big(\frac{z_i-z_j}{h}\Big)\{l(x_i)-l(x_j)\}\nonumber\\
    &=&\frac{1}{hf(z_i)}E\Big\{\mathcal{K}\Big(\frac{z_i-Z}{h}\Big)\{l(x_i)-l(X)\}|z_i\Big\}\nonumber\\
    &=&\frac{1}{hf(z_i)}E\Big\{\mathcal{K}\Big(\frac{z_i-Z}{h}\Big)\{E(l(x_i)|z_i)-E(l(X)|Z)\}\Big\}\nonumber\\
    &=&\frac{1}{f(z_i)}\int\mathcal{K}(u)\{E(l(x_i)|z_i)-E(l(X)|z_i-hu)\}f(z_i-hu)du\nonumber\\
    &=&-h^rk_r\frac{\big(E(l(x_i)|z_i)f(z_i)\big)^{(r)}-E(l(x_i)|z_i)f(z_i)^{(r)}}{f(z_i)}\nonumber\\
    &=:&h^rM(z_i),\label{I_Wg}
\end{eqnarray}
where $u=B^\top(x_i-x)/h$ and Taylor expansion is used in the penultimate step. Also, $(\cdot)^{(r)}$ denotes the $r$th derivative. Thus,
\begin{eqnarray*}
    l(x_i)-\sum_{j=1}^n w_{ij}(\hat B)l(x_j)&=&\sum_{j=1}^n w_{ij}(B)\{l(x_i)-l(x_j)\}+\sum_{j=1}^n [w_{ij}(\hat B)-w_{ij}(B)]\{l(x_i)-l(x_j)\}\nonumber\\
    &=&h^rM(z_i)+o_p(h^r).
\end{eqnarray*}
Together with Lemma~3.3b in Zheng (1996) or Lemma 2 in Guo et al
(2015), it can be easily derived that
\begin{equation}\label{Zhenglemma}
  \frac{1}{n}\sum_{i=1}^n\sum_{k=1}^n w_{ik}(\hat B)\varepsilon_k M(z_i)=O_p(\frac{1}{\sqrt n}).
\end{equation}
Combing the formula (\ref{I_Wg}) and (\ref{Zhenglemma}), we have
\begin{equation*}
  B_{42}=O_p(\frac{h^r}{\sqrt n}).
\end{equation*}
Similarly, we can obtain
\begin{equation*}
  B_{43}=O_p(h^{2r}).
\end{equation*}
Turn to the term $B_{41}$ in (\ref{B4}),
\begin{eqnarray*}
  B_{41}&=&\frac{1}{n}\sum_{i=1}^n\sum_{j=1}^n w_{ij}^2(\hat B)\varepsilon_j^2+
  \frac{1}{n}\sum_{j=1}^n\sum_{k\neq j}^n\varepsilon_j\varepsilon_k\sum_{i=1}^nw_{ij}(\hat B)
  w_{ik}(\hat B)\\
  &=&B_{41,1}+B_{41,2}.
\end{eqnarray*}
As for $B_{41,1}$, we have:
\begin{eqnarray}\label{B411}
  B_{41,1}&=&\frac{1}{n}\sum_{i=1}^n\sum_{j=1}^n\frac{\mathcal{K}^2
  \big((z_i-z_j)/h\big)}{n^2h^2f^2(z_i)}\varepsilon_j^2\nonumber\\
  &=&\frac{1}{n}\sum_{i=1}^n \frac{1}{f^2(z_i)}\times
  \frac{1}{nh^2}E\big\{\big[\varepsilon_j^2\mathcal{K}^2
  \big((Z_i-z_j)/h\big)|Z_i\big]\Big\}\nonumber\\
  &=&\frac{1}{nh}\int \sigma^2(z)dz\int\mathcal{K}^2(u)du+o_p(\frac{1}{n\sqrt h}).
\end{eqnarray}
Thus, the term $B_4$ in (\ref{B4}) can be concluded as
\begin{eqnarray}\label{truB4}
  B_4&=&\frac{1}{n}\sum_{j=1}^n\sum_{k\neq j}^n\varepsilon_j\varepsilon_k\sum_{i=1}^nw_{ij}(\hat B)
  w_{ik}(\hat B)+\frac{1}{nh}\int \sigma^2(z)dz\int\mathcal{K}^2(u)du+O_p(\frac{h^r}{\sqrt n})\nonumber\\
  &&+O_p(h^{2r})+o_p(\frac{1}{n\sqrt h}).
\end{eqnarray}
Recall the term $B_1$ in (\ref{fenzi}),
\begin{eqnarray*}
  B_1&=&\frac{2}{n}\sum_{i=1}^n\varepsilon_i\sum_{j=1}^n w_{ij}(\hat B)[m(x_j)+
  \varepsilon_j-m(x_i)]\\
  &=&\frac{2}{n}\sum_{i=1}^n\varepsilon_i\sum_{j=1}^n w_{ij}(\hat
  B)[m(x_j)-m(x_i)]
  +\frac{2}{n}\sum_{i=1}^n\varepsilon_i^2w_{ii}(\hat B)\\
  &&+\frac{2}{n}\sum_{i=1}^n\sum_{i\neq j}^n\varepsilon_i\varepsilon_jw_{ij}(\hat B)\\
  &=:&B_{11}+B_{12}+B_{13}.
\end{eqnarray*}
Recall the result in (\ref{I_Wg}), it can be easily derived that
\begin{equation*}
  B_{11}=O_p(\frac{h^r}{\sqrt{n}}).
\end{equation*}
For the term $B_{12}$,
\begin{eqnarray}\label{B_12}
  B_{12}&=&\frac{2\mathcal{K}(0)}{nh} E\big\{\varepsilon_i^2\frac{1}{f(z_i)}\big\}
  =\frac{2\mathcal{K}(0)}{nh}\int\sigma^2(z)dz+o_p(\frac{1}{n\sqrt h}).
\end{eqnarray}
Therefore,
\begin{equation}\label{truB1}
  B_1=\frac{2}{n}\sum_{i=1}^n\sum_{i\neq j}^n\varepsilon_i\varepsilon_jw_{ij}(\hat B)+\frac{2\mathcal{K}(0)}{nh}\int\sigma^2(z)dz+o_p(\frac{1}{n\sqrt h})+O_p(\frac{h^r}{\sqrt{n}}).
\end{equation}

Taking formulae (\ref{truB1}),(\ref{truB2}), (\ref{truB3}) and
(\ref{truB4}) into the formula (\ref{fenzi}), we can obtain
\begin{eqnarray}\label{tru_fenzi}
H_n&=&\frac{1}{n}\sum_{i=1}^n\sum_{i\neq j}^n\varepsilon_i\varepsilon_j\big\{w_{ij}(\hat B)
  +w_{ji}(\hat B)-\sum_{k=1}^nw_{ki}(\hat B)w_{kj}(\hat B)\big\}\nonumber\\
  &&+\frac{2Q_1\int\sigma^2(z)f(z)dz}{nh}+
  o_p(\frac{1}{n\sqrt h})\nonumber\\
  &=:&B_{13}-B_{41,2}+\frac{2Q_1\int\sigma^2(z)f(z)dz}{nh}+
  o_p(\frac{1}{n\sqrt h}),
\end{eqnarray}
where
\begin{equation}\label{Q1}
  2Q_1=\Big\{2\mathcal{K}(0)-\int\mathcal{K}^2(u)du\Big\}\frac{\int\sigma^2(z)dz}{
  \int\sigma^2(z)f(z)dz}.
\end{equation}
Considering the term $B_{13}-B_{41,2}$,
\begin{eqnarray}\label{cha_B13411}
  B_{13}-B_{41,2}&=&\frac{1}{n}\sum_{i=1}^n\sum_{i\neq j}^n\varepsilon_i\varepsilon_j\big\{w_{ij}( B)
  +w_{ji}( B)-\sum_{k=1}^nw_{ki}( B)w_{kj}( B)\big\}\nonumber\\
  &&+\frac{1}{n}\sum_{i=1}^n\sum_{i\neq j}^n\varepsilon_i\varepsilon_j\Big\{(w_{ij}(\hat B)-w_{ij}(B))
  +(w_{ji}(\hat B)-w_{ji}(B))\nonumber\\
  &&+\sum_{k=1}^n(w_{ki}(\hat B)w_{kj}(\hat B)-w_{ki}(B)w_{kj}(B))\Big\}\nonumber\\
  &=:&D_1+D_2=D_1+o_p(D_1).
\end{eqnarray}
The last equation holds due to the consistency of $\hat B$ to $B$.

Consequently, $H_n$ can be rewritten as:
\begin{eqnarray}\label{tru_fenzi2}
  H_n&=:&D_1+\frac{2Q_1\int\sigma^2(z)f(z)dz}{nh}+o_p(D_1)+
  o_p(\frac{1}{n\sqrt h}),
\end{eqnarray}

For the term $D_1$ in (\ref{tru_fenzi2}), based on Whittle (1964),
Jong (1987) and Dette (2000), we can similarly prove that
\begin{equation*}
  n\sqrt{h}\Big(H_n-\frac{2Q_1\int\sigma^2(z)f(z)dz}{nh}\Big)\Rightarrow N(0,\varepsilon_0^2).
\end{equation*}
Recall the expression of statistic $T_n$ in (\ref{statistic}) and Lemma~\ref{lema1}, according to Slutsky's Theorem, we can obtain
\begin{equation*}
  \sqrt{h}\Big(T_n-\frac{Q_1}{h}\Big)=\frac{n\sqrt{h}\Big(H_n-
  (2Q_1/nh)\int\sigma^2(z)f(z)dz\Big)}{2RSS_1/n}
  \Rightarrow N(0,V_0),
\end{equation*}
where
\begin{equation}\label{V0}
  V_0=\frac{\eta_0^2}{4(\int\sigma^2(z)f(z)dz)^2}
  =\frac{\int\sigma^4(z)dz\int[2\mathcal{K}(u)-
   \mathcal{K}*\mathcal{K}(u)]^2du}{2(\int\sigma^2(z)f(z)dz)^2},
\end{equation}
and $\mathcal{K}*\mathcal{K}$ denotes the convolution of
$\mathcal{K}$ and $\mathcal{K}$.

The proof of Theorem~\ref{theo1} is finished. $\Box$

\vspace{0.5cm} \noindent \textit{Proof of Theorem~\ref{theo_nullnobias}.}
Under the null hypothesis in (\ref{null_model}),
$m(x)=g(\beta^\top x,\theta)$ holds and from Lemma~\ref{lema4},
we have $\widetilde{\mbox{RSS}}_1/n=\sum_{i=1}^n [y_i-\hat {\tilde m}(\hat B(\hat q)^\top x_i)][y_i-g(\hat\beta^\top x_i,\hat\theta)]/n+o_p(1)$. As for the numerator of the statistic $\tilde{T}_n$ in (\ref{nobias_statistic}), denote $J_n=(\mbox{RSS}_0-\widetilde{\mbox{RSS}}_1)/n$, then we have
\begin{eqnarray*}
 J_n&=&\frac{1}{n}\sum_{i=1}^n\big[\hat {\tilde m}(\hat B(\hat q)^\top x_i)-g(\hat\beta^\top x_i,\hat\theta)\big]\big[y_i-g(\hat\beta^\top x_i,\hat\theta)\big]+o_p(1)\\
 &=&\frac{1}{n}\sum_{i=1}^n \Big[\sum_{j\neq i}^n \tilde{w}_{ij}(\hat B) m(x_j)-m(B^\top x_i)+g(\beta^\top x_i,\theta)-g(\hat\beta^\top x_i,\hat\theta)+\sum_{j\neq i}^n \tilde{w}_{ij}(\hat B)\varepsilon_j\Big]\\
 &&\times [g(\beta^\top x_i,\theta)+\varepsilon_i-g(\hat\beta^\top x_i,\hat\theta)]+o_p(1)\\
 &=&\frac{1}{n}\sum_{i=1}^n \Big[\sum_{j\neq i}^n \tilde{w}_{ij}(\hat B) m(x_j)-m(B^\top x_i)\Big] [g(\beta^\top x_i,\theta)+\varepsilon_i-g(\hat\beta^\top x_i,\hat\theta)]\\
 &&+\frac{1}{n}\sum_{i=1}^n [g(\beta^\top x_i,\theta)-g(\hat\beta^\top x_i,\hat\theta)]
 [g(\beta^\top x_i,\theta)+\varepsilon_i-g(\hat\beta^\top x_i,\hat\theta)]\\
 &&+\frac{1}{n}\sum_{i=1}^n  [g(\beta^\top x_i,\theta)+\varepsilon_i-g(\hat\beta^\top x_i,\hat\theta)]\sum_{j\neq i}^n \tilde{w}_{ij}(\hat B) \varepsilon_j+o_p(1)\\
 &=:&J_{n1}+J_{n2}+J_{n3}+o_p(1).
\end{eqnarray*}

For the first term $J_{n1}$, similarly as the proof of (\ref{I_Wg}), it is not difficult to obtain that $\sum_{j\neq i}^n \tilde{w}_{ij}(\hat B) m(x_j)-m(B^\top x_i)=O_p(h^r)$, further we have
\begin{eqnarray*}
  J_{n1}&=&\frac{1}{n}\sum_{i=1}^n \Big[\sum_{j\neq i}^n \tilde{w}_{ij}(\hat B) m(x_j)-m(B^\top x_i)\Big] [g(\beta^\top x_i,\theta)-g(\hat\beta^\top x_i,\hat\theta)]\\
  &&+\frac{1}{n}\sum_{i=1}^n \varepsilon_i\Big[\sum_{j\neq i}^n \tilde{w}_{ij}(\hat B) m( x_j)-m(B^\top x_i)\Big]\\
  &=&O_p(\frac{1}{\sqrt{n}})\times O_p(h^r)\times O_p(\frac{1}{\sqrt{n}})+O_p(\frac{1}{\sqrt{n}})\times O_p(h^r)\\
  &=&O_p(\frac{h^r}{\sqrt{n}}).
\end{eqnarray*}

As to the second term $J_{n2}$, just the same as the terms $B_2$ in (\ref{truB2}) and $B_3$ in (\ref{truB3}), we can obtain $J_{n2}=O_p(1/n)$.

Turn to the term $J_{n3}$, we first define the following term:
\begin{eqnarray*}
  J_{n3}^\star&=&\frac{1}{n(n-1)}\sum_{i=1}^n\sum_{j\neq i}^n\frac{\mathcal{K}_h\{B^\top(x_i-x_j)\}\varepsilon_jg'(\tilde{\alpha}_2)
  (\alpha_0-\hat\alpha)}{f(x_i)}\\
  &&+\frac{1}{n(n-1)}\sum_{i=1}^n\sum_{j\neq i}^n\frac{\mathcal{K}_h\{B^\top(x_i-x_j)\}\varepsilon_i\varepsilon_j}{f(x_i)}\\
  &=&J_{n3,1}^\star+J_{n3,2}^\star,
\end{eqnarray*}
where $\mathcal{K}_h(\cdot)=\mathcal{K}(\cdot/h)/h$ and $\tilde{\alpha}_2$ lies between $\alpha_0$ and $\hat\alpha$. It is not difficult to derive that $J_{n3}=J_{n3}^\star+o_p(J_{n3}^\star)$. As to the term $J_{n3,1}^\star$, we can conclude that $J_{n3,1}^\star=o_p(1/n)$. Note that $J_{n3,2}^\star$ is a degenerate U-statistic. Following Zheng (1996), it can be easily obtained that $nh^{1/2}J_{n3,2}^\star\Rightarrow N(0,\Sigma_1)$ with $\Sigma_1=2\int \mathcal{K}^2(u)du\int\sigma^4(z)dz$. Under the condition $nh^{1/2+2r}\Rightarrow 0$, we have $nh^{1/2}J_{n1}\Rightarrow 0$. Further it can be concluded that $nh^{1/2}J_n\Rightarrow N(0,\Sigma_1)$. Based on Lemma~\ref{lema4} and through Slutsky's Theorem, Theorem~\ref{theo_nullnobias} can be obtained.
The proof of Theorem~\ref{theo_nullnobias} is completed. $\Box$

\vspace{0.5cm} \noindent \textit{Proof of Theorem~\ref{theo2}.} We
first consider the global alternative (\ref{nonpara_model}).  Under
this alternative, Remark~\ref{consistency_q} shows that $\hat
q\rightarrow q\geq 1$. From Lemma~\ref{lema2}, $\hat\alpha$ is  a
root-$n$ consistent estimate of $\tilde{\alpha}$ which is different
from the true value $\alpha_0$ under the null hypothesis. Let
$\Delta({x}_i)= m(B^\top x_i)-g(\tilde{\beta}^\top
x_i,\tilde\theta)$. Then
$y_i=\varepsilon_i+m(x_i)=\varepsilon_i+\Delta({x}_i)+g(\tilde{\beta}^\top
{x}_i,\tilde\theta)$. Similar to the proof of Theorem 1, it is then
easy to see that $H_n=(RSS_0-RSS_1)/n\Rightarrow E(\Delta^2({X}))$.
Together with Lemma~\ref{lema1}, we can obtain that $T_n/n\to
Constant>0$ in probability. Thus $\sqrt{h}T_n=\sqrt{h}n\times
T_n/n\rightarrow\infty.$

Under the local alternative hypothesis in (\ref{loca_alter}), also
denote $H_n=(\mbox{RSS}_0-\mbox{RSS}_1)/n$ and we have
\begin{eqnarray}\label{Hn_decom}
   H_n&=& \frac{1}{n}\sum_{i=1}^n [y_i-g(\hat\beta^\top x_i,\hat\theta)]^2-\frac{1}{n}\sum_{i=1}^n[y_i-\hat m (\hat B^\top x_i)]^2\nonumber\\
  &=&\frac{1}{n}\sum_{i=1}^n \big[2y_i-g(\hat\beta^\top x_i,\hat\theta)-\hat m (\hat B^\top x_i)\big]\big[\hat m (\hat B^\top x_i)-g(\hat\beta^\top x_i,\hat\theta)\big]\nonumber\\
  &=&\frac{1}{n}\sum_{i=1}^n\big[2\varepsilon_i+m(x_i)-\hat m (\hat B^\top x_i)+g(\tilde\beta^\top x_i,\tilde\theta)+C_n m(B^\top x_i)-g(\hat\beta^\top x_i,\hat\theta)\big]\nonumber\\
  &&\big[\hat m (\hat B^\top x_i)-m(x_i)+g(\tilde\beta^\top x_i,\tilde\theta)+C_n m(B^\top x_i)-g(\hat\beta^\top x_i,\hat\theta)\big]\nonumber\\
  &=&\frac{2}{n}\sum_{i=1}^n\varepsilon_i[\hat m (\hat B^\top x_i)-m(x_i)]+\frac{2}{n}\sum_{i=1}^n\varepsilon_i[g(\tilde\beta^\top x_i,\tilde\theta)-g(\hat\beta^\top x_i,\hat\theta)]+\frac{2C_n}{n}\sum_{i=1}^n\varepsilon_im(B^\top x_i)\nonumber\\
  &&+\frac{1}{n}\sum_{i=1}^n[g(\tilde\beta^\top x_i,\tilde\theta)+C_n m(B^\top x_i)-g(\hat\beta^\top x_i,\hat\theta)]^2-\frac{1}{n}\sum_{i=1}^n[\hat m (\hat B^\top x_i)-m(x_i)]^2\nonumber\\
  &=:&H_{n1}+H_{n2}+H_{n3}+H_{n4}-H_{n5}.
\end{eqnarray}

From Lemma~\ref{lema2}, $\hat\alpha-\tilde\alpha=O_p(1/\sqrt n)$.
Thus similar to the derivation of $B_2$, it is not difficult to
verify that
\begin{equation*}
  n\sqrt{h}H_{n2}=n\sqrt{h}\times O_p(\frac{1}{n})=o_p(1).
\end{equation*}

For $H_{n3}$, since $E\{\varepsilon m(B^\top X)\}=0$ and $Var\{\varepsilon m(B^\top X)\}\rightarrow c$, where $c$ is a constant, when $C_n=n^{-1/2}h^{-1/4}$, we have
\begin{equation*}
 n\sqrt{h} H_{n3}=n\sqrt{h}\times O_p(\frac{C_n}{\sqrt n})=o_p(1).
\end{equation*}

Turn to the term $H_{n4}$,
\begin{eqnarray*}
  H_{n4}&=&\frac{C_n^2}{n}\sum_{i=1}^n m^2(B^\top x_i)-\frac{2C_n}{n}\sum_{i=1}^n m(B^\top x_i)[g(\hat\beta^\top x_i,\hat\theta)-g(\tilde\beta^\top x_i,\tilde\theta)]\\
  &&+\frac{1}{n}\sum_{i=1}^n[g(\hat\beta^\top x_i,\hat\theta)-g(\tilde\beta^\top x_i,\tilde\theta)]^2\\
  &=&H_{n4,1}-2H_{n4,2}+H_{n4,3}.
\end{eqnarray*}
For the term $H_{n4,1}$, when $C_n=n^{-1/2}h^{-1/4}$,
\begin{equation*}
  n\sqrt{h}H_{n4,1}=n\sqrt{h}C_n^2 \times E[m^2(B^\top X)]+o_p(1)=E[m^2(B^\top X)]+o_p(1).
\end{equation*}
With regard to the term $H_{n4,2}$,
\begin{equation*}
  n\sqrt{h}H_{n4,2}=n\sqrt{h}\times O_p(\frac{C_n}{\sqrt n})=o_p(1).
\end{equation*}
Similar to the formula (\ref{truB3}), we have
\begin{equation*}
  n\sqrt{h}H_{n4,3}=n\sqrt{h}\times O_p(\frac{1}{n})=o_p(1).
\end{equation*}
According to the above three formulae, we can derive that
\begin{equation*}
  n\sqrt{h}H_{n4}=E[m^2(B^\top X)]+o_p(1).
\end{equation*}

It is easy to verify that the terms $H_{n1}$ and $H_{n5}$ in
(\ref{Hn_decom}) are the same as the terms $B_1$ and $B_4$ in
(\ref{fenzi}), respectively. Thus similar to the proof of
Theorem~{\ref{theo1}}, under the local alternative hypothesis
(\ref{loca_alter}), we can derive that
\begin{equation*}
  \sqrt h (T_n-\frac{Q_1}{ h})\Rightarrow N(\mu_1,V_0),
\end{equation*}
where
\begin{equation*}
  \mu_1=\frac{E[m^2(B^\top X)]}{2\int \sigma^2(z)f(z)dz},
\end{equation*}
and $V_0$ has been defined in (\ref{V0}).

Theorem~\ref{theo2} is proved. $\Box$

\vspace{0.5cm} \noindent \textit{Proof of Theorem~\ref{theo_alternobias}.}
Consider the global alternative in (\ref{nonpara_model}) first. Denote $\Delta(x_i)=m(B^\top x_i)-g(\tilde{\beta}^\top x_i,\tilde\theta)$, thus, $y_i=m(x_i)+\varepsilon_i=\Delta(x_i)+g(\tilde{\beta}^\top x_i,\tilde\theta)+\varepsilon_i$.
We first consider the denominator $\widetilde{\mbox{RSS}}_1$ in the statistic $\tilde{T}_n$ in (\ref{nobias_statistic}). Based on the formula (\ref{ymyg}), it can be obtained that
\begin{equation*}
  \frac{1}{n}\widetilde{\mbox{RSS}}_1=\frac{1}{n}\sum_{i=1}^n|\varepsilon_i^2+\varepsilon_i\Delta(x_i)|+o_p(1).
\end{equation*}
Note the conditions in Appendix, suppose $E|\varepsilon|<\infty$, $E|g(\beta^\top X,\theta)|<\infty$ and $E|m(B^\top X)|<\infty$, we have
\begin{equation}\label{RSS1k1}
  \widetilde{\mbox{RSS}}_1/n\Rightarrow K_1>0,
\end{equation}
where $K_1$ is a positive constant. Turn to the numerator of statistic $\tilde{T}_n$, denote $J_n=(\mbox{RSS}_0-\widetilde{\mbox{RSS}}_1)/n$. In order to remove the absolute value sign of $\widetilde{\mbox{RSS}}_1$, we consider every term in it as follows.
For $i=1,2,\ldots,n$, assume that there are $m(m\leq n)$ terms which satisfy $[y_i-\hat {\tilde m}(\hat B(\hat q)^\top x_i)][y_i-g(\hat\beta^\top x_i,\hat\theta)]>0$. As to these $m$ terms,
we have
\begin{eqnarray}
  &&\frac{1}{n}\{\mbox{RSS}_0-\widetilde{\mbox{RSS}}_1\}_{\{m\}}\nonumber\\
  &=:&\frac{1}{n}\sum_{\{m\}}[y_i-g(\hat\beta^\top x_i,\hat\theta)]\Big[\hat {\tilde m}(\hat B(\hat q)^\top x_i)-g(\hat\beta^\top x_i,\hat\theta)\Big]+o_p(1)\nonumber\\
  &=&\frac{1}{n}\sum_{\{m\}}[\Delta(x_i)+\varepsilon_i+g(\tilde{\beta}^\top x_i,\tilde\theta)-g(\hat\beta^\top x_i,\hat\theta)]\nonumber\\
  &&\times\big[\hat {\tilde m}(\hat B(\hat q)^\top x_i)-m(B^\top x_i)+\Delta(x_i)+g(\tilde{\beta}^\top x_i,\tilde\theta)-g(\hat\beta^\top x_i,\hat\theta)\big]+o_p(1)\nonumber\\
  &=&\frac{1}{n}\sum_{\{m\}}\Delta^2(x_i)+o_p(1).\label{J_n1glo}
\end{eqnarray}
For another $n-m$ terms which have $[y_i-\hat {\tilde m}(\hat B(\hat q)^\top x_i)][y_i-g(\hat\beta^\top x_i,\hat\theta)]<0$, thus,
\begin{eqnarray}
  &&\frac{1}{n}\{\mbox{RSS}_0-\widetilde{\mbox{RSS}}_1\}_{\{n-m\}}\nonumber\\
  &=:&\frac{1}{n}\sum_{\{n-m\}}[y_i-g(\hat\beta^\top x_i,\hat\theta)]\Big[2y_i-\hat {\tilde m}(\hat B(\hat q)^\top x_i)-g(\hat\beta^\top x_i,\hat\theta)\Big]+o_p(1)\nonumber\\
  &=&\frac{1}{n}\sum_{\{m\}}[\Delta(x_i)+\varepsilon_i+g(\tilde{\beta}^\top x_i,\tilde\theta)-g(\hat\beta^\top x_i,\hat\theta)]\nonumber\\
  &&\times\Big[m(B^\top x_i)-\hat {\tilde m}(\hat B(\hat q)^\top x_i)+\Delta(x_i)+g(\tilde{\beta}^\top x_i,\tilde\theta)-g(\hat\beta^\top x_i,\hat\theta)+2\varepsilon_i\Big]+o_p(1)\nonumber\\
  &=&\frac{1}{n}\sum_{\{n-m\}}[\Delta^2(x_i)+2\varepsilon_i^2]+o_p(1).\label{J_n2glo}
\end{eqnarray}
Combining (\ref{J_n1glo}) and (\ref{J_n2glo}), under the conditions in Appendix, it can be derived that
\begin{equation}\label{J_n}
  J_n=\frac{1}{n}\{\mbox{RSS}_0-\widetilde{\mbox{RSS}}_1\}_{\{m\}}+
  \frac{1}{n}\{\mbox{RSS}_0-\widetilde{\mbox{RSS}}_1\}_{\{n-m\}}\Rightarrow K_2,
\end{equation}
where $K_2$ is a positive constant. Based on (\ref{RSS1k1}) and (\ref{J_n}), we can obtain that
$\tilde{T}_n/n=J_n/(\widetilde{\mbox{RSS}}_1/n)\rightarrow K_2/K_1=C_1>0$ in probability, where $C_1$ is a positive constant. Further $\sqrt{h}\tilde{T}_n=n\sqrt{h}\times\tilde{T}_n/n\rightarrow\infty$, which completes the proof under global alternative hypothesis.

\vspace{0.3cm}
Under the local alternative hypothesis in (\ref{loca_alter}),  from Lemma~\ref{lema4},
we have $\widetilde{\mbox{RSS}}_1/n=\sum_{i=1}^n [y_i-\hat {\tilde m}(\hat B(\hat q)^\top x_i)][y_i-g(\hat\beta^\top x_i,\hat\theta)]/n+o_p(1)$. Let $J_n=(\mbox{RSS}_0-\widetilde{\mbox{RSS}}_1)/n$ and we have
\begin{eqnarray*}
 J_n&=&\frac{1}{n}\sum_{i=1}^n\big[\hat {\tilde m}(\hat B(\hat q)^\top x_i)-g(\hat\beta^\top x_i,\hat\theta)\big]\big[y_i-g(\hat\beta^\top x_i,\hat\theta)\big]+o_p(1)\\
 &=&\frac{1}{n}\sum_{i=1}^n \Big[\sum_{j\neq i}^n \tilde{w}_{ij}(\hat B) \{m(x_j)+\varepsilon_j\}-m(B^\top x_i)+g(\tilde{\beta}^\top x_i,\tilde{\theta})-g(\hat\beta^\top x_i,\hat\theta)+C_nm(B^\top x_i)\Big]\\
 &&\times [g(\tilde{\beta}^\top x_i,\tilde{\theta})+\varepsilon_i+C_n m(B^\top x_i)-g(\hat\beta^\top x_i,\hat\theta)]+o_p(1)\\
 &=&\frac{1}{n}\sum_{i=1}^n \Big[\sum_{j\neq i}^n \tilde{w}_{ij}(\hat B) \{m(x_j)+\varepsilon_j\}-m(B^\top x_i)+g(\tilde{\beta}^\top x_i,\tilde{\theta})-g(\hat\beta^\top x_i,\hat\theta)\Big]\\
  &&\times[g(\tilde{\beta}^\top x_i,\tilde{\theta})+\varepsilon_i-g(\hat\beta^\top x_i,\hat\theta)]+\frac{C_n}{n}\sum_{i=1}^nm(B^\top x_i)[g(\tilde{\beta}^\top x_i,\tilde{\theta})+\varepsilon_i-g(\hat\beta^\top x_i,\hat\theta)]\\
  &&+\frac{C_n}{n}\sum_{i=1}^nm(B^\top x_i)\Big[\sum_{j\neq i}^n \tilde{w}_{ij}(\hat B) \{m(x_j)+\varepsilon_j\}-m(B^\top x_i)+g(\tilde{\beta}^\top x_i,\tilde{\theta})-g(\hat\beta^\top x_i,\hat\theta)\Big]\\
  &&+\frac{C_n^2}{n}\sum_{i=1}^nm^2(B^\top x_i)+o_p(1)\\
&=:&I_{n1}+I_{n2}+I_{n3}+I_{n4}+o_p(1).
\end{eqnarray*}
For the term $I_{n1}$, from the proof of Theorem~\ref{theo_nullnobias}, it can be obtained that $nh^{1/2}I_{n1}\Rightarrow N(0,\Sigma_1)$. Due to the fact that $E(\varepsilon|X)=0$ and the root-$n$ consistency of $\hat\alpha=(\hat\beta,\hat\theta)^\top$ to $\tilde\alpha=(\tilde\beta,\tilde\theta)^\top$, it can be not difficult shown that $I_{n2}=O_p(C_n/\sqrt{n})$. Thus, when $C_n=n^{-1/2}h^{-1/4}$, $nh^{1/2}I_{n2}\Rightarrow 0$. As to the term $I_{n4}$ with $C_n=n^{-1/2}h^{-1/4}$, we have $nh^{1/2}I_{n4}\Rightarrow E\{m^2(B^\top X)\}$. Finally, consider the term $I_{n3}$,
\begin{eqnarray*}
  I_{n3}&=&\frac{C_n}{n}\sum_{i=1}^nm(B^\top x_i)\sum_{j\neq i}^n \tilde{w}_{ij}(\hat B) \varepsilon_j+\frac{C_n}{n}\sum_{i=1}^nm(B^\top x_i)\Big[\sum_{j\neq i}^n \tilde{w}_{ij}(\hat B) m(x_j)-m(B^\top x_i)\Big]\\
  &&+\frac{C_n}{n}\sum_{i=1}^nm(B^\top x_i)[g(\tilde{\beta}^\top x_i,\tilde{\theta})-g(\hat\beta^\top x_i,\hat\theta)]\\
  &=:&I_{n3,1}+I_{n3,2}+I_{n3,3}.
\end{eqnarray*}
From Zheng (1996), it can be known that $I_{n3,1}=O_p(C_n/\sqrt{n})$.
For $I_{n3,2}$, from the proof of Theorem~\ref{theo_nullnobias}, it can be derived that
$I_{n3,2}=O_p(C_n h^r)$. Finally due to the fact that $\hat\alpha=(\hat\beta,\hat{\theta})^\top$
is root-$n$ consistent estimate of $\tilde\alpha=(\tilde\beta,\tilde{\theta})^\top$, we can
conclude that $I_{n3,3}=O_p(C_n/\sqrt{n})$. With
$C_n=n^{-1/2}h^{-1/4}$ and condition (C5), we can get that
$nh^{1/2}I_{n3}=o_p(1)$. Thus based on Lemma~\ref{lema4} and using Slutsky's
Theorem, Theorem~\ref{theo_alternobias} is obtained.

The proof of Theorem~\ref{theo_alternobias} is finished. $\Box$

\newpage
\begin{table}[h!]
\caption{\linespread{1.15}\small Empirical sizes and powers of $\tilde{S}_n^{OPG},\tilde{R}_n^{OPG}$ and $\tilde{S}_n^{MAVE},\tilde{R}_n^{MAVE}$ for $H_0$ v.s. $H_{11}$ and $H_{12}$ at significance level $\alpha=0.05$ with $p=8$.
}
\footnotesize
\begin{center}
\begin{tabular}{ccccccccccccccccccc}
\hline
\multicolumn{1}{c}{\multirow{2}{*}{}}&
\multicolumn{1}{c}{\multirow{2}{*}{$\varepsilon$}}&
\multicolumn{1}{c}{\multirow{2}{*}{$a_{\cdot}$}}&
\multicolumn{4}{c}{$n=100$}& &\multicolumn{4}{c}{$n=200$}
 \\
\cline{4-7} \cline{9-12}
\multicolumn{1}{c}{}&\multicolumn{1}{c}{}&\multicolumn{1}{c}{}&
\multicolumn{1}{c}{$\tilde{S}_n^{OPG}$}
&\multicolumn{1}{c}{$\tilde{R}_n^{OPG}$}&
\multicolumn{1}{c}{$\tilde{S}_n^{MAVE}$}&
\multicolumn{1}{c}{$\tilde{R}_n^{MAVE}$}&
&
\multicolumn{1}{c}{$\tilde{S}_n^{OPG}$}
&\multicolumn{1}{c}{$\tilde{R}_n^{OPG}$}&
\multicolumn{1}{c}{$\tilde{S}_n^{MAVE}$}&
\multicolumn{1}{c}{$\tilde{R}_n^{MAVE}$}
\\
\hline
$H_{11}$&$\varepsilon\sim N(0,1)$
&0&0.045&0.047&0.046&0.051&&0.043&0.051&0.043&0.052\\
&&0.1&0.062&0.069&0.055&0.064&&0.072&0.085&0.080&0.086\\
&&0.2&0.160&0.175&0.175&0.179&&0.314&0.340&0.308&0.343\\
&&0.3&0.376&0.428&0.416&0.439&&0.726&0.762&0.712&0.756\\
&&0.4&0.695&0.728&0.712&0.731&&0.966&0.971&0.967&0.979\\
&&0.5&0.899&0.917&0.909&0.918&&0.998&0.999&1.000&1.000\\
\hline
&$\varepsilon\sim t(5)$
&0&0.060&0.045&0.065&0.046&&0.059&0.048&0.059&0.046\\
&&0.1&0.077&0.073&0.079&0.075&&0.083&0.077&0.086&0.084\\
&&0.2&0.143&0.141&0.144&0.142&&0.211&0.216&0.220&0.226\\
&&0.3&0.268&0.265&0.276&0.287&&0.467&0.512&0.449&0.510\\
&&0.4&0.449&0.474&0.486&0.489&&0.780&0.817&0.790&0.807\\
&&0.5&0.697&0.728&0.700&0.721&&0.949&0.966&0.950&0.976\\
\hline
$H_{12}$&$\varepsilon\sim N(0,1)$
&0&0.053&0.049&0.040&0.047&&0.046&0.048&0.055&0.053\\
&&0.1&0.132&0.135&0.128&0.125&&0.213&0.232&0.221&0.217\\
&&0.2&0.532&0.545&0.548&0.543&&0.879&0.887&0.870&0.866\\
&&0.3&0.918&0.928&0.916&0.915&&0.998&0.999&0.999&0.999\\
&&0.4&0.996&0.998&0.995&0.996&&1.000&1.000&1.000&1.000\\
&&0.5&1.000&1.000&0.999&1.000&&1.000&1.000&1.000&1.000\\
\hline
&$\varepsilon\sim t(5)$
&0&0.061&0.054&0.078&0.055&&0.061&0.055&0.063&0.048\\
&&0.1&0.117&0.100&0.126&0.103&&0.147&0.149&0.156&0.152\\
&&0.2&0.385&0.377&0.376&0.371&&0.590&0.638&0.624&0.649\\
&&0.3&0.717&0.727&0.730&0.727&&0.958&0.966&0.964&0.965\\
&&0.4&0.943&0.948&0.937&0.938&&0.999&0.999&0.998&0.999\\
&&0.5&0.992&0.992&0.993&0.999&&1.000&1.000&1.000&1.000\\
\hline
\end{tabular}\label{study1_1}
\end{center}
\end{table}

\begin{table}[h!]
\caption{\linespread{1.15}\small Empirical sizes and powers of $\tilde{S}_n^{OPG},\tilde{R}_n^{OPG}$ and $\tilde{S}_n^{MAVE},\tilde{R}_n^{MAVE}$ for $H_0$ v.s. $H_{13}$ and $H_{14}$ at significance level $\alpha=0.05$ with $p=8$.
}
\footnotesize
\begin{center}
\begin{tabular}{ccccccccccccccccccc}
\hline
\multicolumn{1}{c}{\multirow{2}{*}{}}&
\multicolumn{1}{c}{\multirow{2}{*}{$\varepsilon$}}&
\multicolumn{1}{c}{\multirow{2}{*}{$a_{\cdot}$}}&
\multicolumn{4}{c}{$n=100$}& &\multicolumn{4}{c}{$n=200$}
 \\
\cline{4-7} \cline{9-12}
\multicolumn{1}{c}{}&\multicolumn{1}{c}{}&\multicolumn{1}{c}{}&
\multicolumn{1}{c}{$\tilde{S}_n^{OPG}$}
&\multicolumn{1}{c}{$\tilde{R}_n^{OPG}$}&
\multicolumn{1}{c}{$\tilde{S}_n^{MAVE}$}&
\multicolumn{1}{c}{$\tilde{R}_n^{MAVE}$}&
&
\multicolumn{1}{c}{$\tilde{S}_n^{OPG}$}
&\multicolumn{1}{c}{$\tilde{R}_n^{OPG}$}&
\multicolumn{1}{c}{$\tilde{S}_n^{MAVE}$}&
\multicolumn{1}{c}{$\tilde{R}_n^{MAVE}$}
\\
\hline
$H_{13}$&$\varepsilon\sim N(0,1)$
&0&0.043&0.048&0.042&0.046&&0.036&0.046&0.042&0.053\\
&&0.2&0.088&0.090&0.091&0.089&&0.144&0.146&0.131&0.139\\
&&0.4&0.300&0.308&0.305&0.297&&0.617&0.635&0.636&0.648\\
&&0.6&0.677&0.683&0.697&0.694&&0.970&0.974&0.974&0.977\\
&&0.8&0.927&0.928&0.921&0.922&&0.999&0.999&0.999&1.000\\
&&1.0&0.989&0.990&0.992&0.993&&1.000&1.000&1.000&1.000\\
\hline
&$\varepsilon\sim t(5)$
&0&0.062&0.051&0.071&0.054&&0.055&0.047&0.061&0.047\\
&&0.2&0.090&0.087&0.098&0.087&&0.106&0.094&0.122&0.107\\
&&0.4&0.194&0.191&0.218&0.216&&0.375&0.392&0.402&0.417\\
&&0.6&0.445&0.445&0.458&0.458&&0.800&0.813&0.794&0.830\\
&&0.8&0.700&0.710&0.727&0.735&&0.962&0.968&0.975&0.979\\
&&1.0&0.892&0.901&0.889&0.902&&0.998&0.997&0.996&0.997\\
\hline
$H_{14}$&$\varepsilon\sim N(0,1)$
&0&0.060&0.049&0.058&0.048&&0.059&0.054&0.045&0.053\\
&&0.2&0.068&0.066&0.074&0.082&&0.090&0.107&0.092&0.106\\
&&0.4&0.201&0.210&0.197&0.212&&0.416&0.446&0.443&0.458\\
&&0.6&0.448&0.488&0.500&0.501&&0.885&0.879&0.895&0.892\\
&&0.8&0.768&0.769&0.797&0.798&&0.993&0.988&0.997&0.995\\
&&1.0&0.920&0.931&0.946&0.943&&1.000&1.000&1.000&1.000\\
\hline
&$\varepsilon\sim t(5)$
&0&0.081&0.046&0.070&0.045&&0.073&0.049&0.064&0.054\\
&&0.2&0.103&0.063&0.104&0.075&&0.090&0.078&0.090&0.071\\
&&0.4&0.152&0.153&0.164&0.158&&0.240&0.246&0.261&0.265\\
&&0.6&0.309&0.318&0.318&0.330&&0.598&0.643&0.640&0.666\\
&&0.8&0.507&0.534&0.534&0.556&&0.893&0.900&0.922&0.929\\
&&1.0&0.719&0.759&0.761&0.782&&0.987&0.987&0.989&0.990\\
\hline
\end{tabular}\label{study1_2}
\end{center}
\end{table}

\begin{table}[h!]
\caption{\linespread{1.15}\small Empirical sizes and powers of $\tilde{R}_n^{OPG}$ and $T_{n,A}^{FZZ},T_{n,B}^{FZZ}$ for $H_0$ v.s. $H_{21}$  at significance level $\alpha=0.05$ with $p=4$.
}
\footnotesize
\begin{center}
\begin{tabular}{ccccccccccccccccccc}
\hline
\multicolumn{1}{c}{\multirow{2}{*}{}}&
\multicolumn{1}{c}{\multirow{2}{*}{$X$}}&
\multicolumn{1}{c}{\multirow{2}{*}{$a_{\cdot}$}}&
\multicolumn{3}{c}{$n=100$}& &\multicolumn{3}{c}{$n=200$}
 \\
\cline{4-6} \cline{8-10}
\multicolumn{1}{c}{}&\multicolumn{1}{c}{}&\multicolumn{1}{c}{}&
\multicolumn{1}{c}{$\tilde{R}_n^{OPG}$}
&\multicolumn{1}{c}{$T_{n,A}^{FZZ}$}&
\multicolumn{1}{c}{$T_{n,B}^{FZZ}$}&
&
\multicolumn{1}{c}{$\tilde{R}_n^{OPG}$}
&\multicolumn{1}{c}{$T_{n,A}^{FZZ}$}&
\multicolumn{1}{c}{$T_{n,B}^{FZZ}$}
\\
\hline
$H_{21}$&$X\sim N(0,\Sigma_1)$
&0&0.054&0.069&0.051&&0.051&0.067&0.046\\
&&0.2&0.096&0.073&0.061&&0.111&0.068&0.061\\
&&0.4&0.234&0.074&0.086&&0.387&0.069&0.119\\
&&0.6&0.451&0.074&0.143&&0.778&0.077&0.223\\
&&0.8&0.729&0.082&0.247&&0.968&0.078&0.435\\
&&1.0&0.882&0.083&0.359&&0.998&0.079&0.666\\
\hline
&$X\sim N(0,\Sigma_2)$
&0&0.050&0.077&0.049&&0.046&0.068&0.050\\
&&0.2&0.086&0.078&0.062&&0.110&0.072&0.078\\
&&0.4&0.246&0.082&0.117&&0.502&0.073&0.144\\
&&0.6&0.575&0.083&0.193&&0.892&0.075&0.354\\
&&0.8&0.818&0.085&0.334&&0.987&0.076&0.665\\
&&1.0&0.946&0.089&0.510&&1.000&0.077&0.878\\
\hline
$H_{22}$&$X\sim N(0,\Sigma_1)$
&0&0.049&0.067&0.048&&0.048&0.071&0.047\\
&&0.3&0.139&0.073&0.059&&0.210&0.072&0.058\\
&&0.6&0.474&0.074&0.074&&0.750&0.073&0.086\\
&&0.9&0.818&0.077&0.094&&0.991&0.074&0.133\\
&&1.2&0.957&0.078&0.147&&0.999&0.076&0.252\\
&&1.5&0.993&0.083&0.221&&1.000&0.083&0.443\\
\hline
&$X\sim N(0,\Sigma_2)$
&0&0.051&0.077&0.048&&0.048&0.068&0.047\\
&&0.3&0.076&0.084&0.065&&0.102&0.070&0.048\\
&&0.6&0.210&0.085&0.073&&0.363&0.072&0.099\\
&&0.9&0.457&0.086&0.113&&0.782&0.075&0.156\\
&&1.2&0.754&0.089&0.177&&0.987&0.077&0.300\\
&&1.5&0.925&0.090&0.265&&1.000&0.078&0.484\\
\hline
\end{tabular}\label{study2}
\end{center}
\end{table}

\newpage

\newpage

\newpage

\newpage

\end{document}